\documentclass[journal]{IEEEtran}

\usepackage{times,amsmath,epsfig}
\usepackage{mathtools,nccmath}
\usepackage{amsfonts,dsfont}
\usepackage{amscd}
\usepackage{amssymb}
\usepackage{indentfirst}
\usepackage{stfloats}
\usepackage{color}
\usepackage{hyperref}
\usepackage[noadjust]{cite}
\usepackage{nicefrac}
\usepackage{multicol}
\usepackage{multirow}
\usepackage{booktabs}

\usepackage[linesnumbered, ruled, lined]{algorithm2e}
\usepackage{graphicx}
\DeclareGraphicsExtensions{.pdf,.jpeg,.png}
\usepackage{siunitx} 
\usepackage[caption=false]{subfig}

\begin{document}

\title{An iALM-ICA-based Anti-Jamming DS-CDMA Receiver for LMS Systems}

\author{Hyoyoung~Jung, \textit{Student~Member,~IEEE},
	 Jaewook~Kang, \textit{Member,~IEEE},
	 Tae~Seok~Lee, \textit{Student~Member,~IEEE},
	 Suil~Kim, \textit{Member,~IEEE},
	 and Kiseon~Kim, \textit{Senior~Member,~IEEE}

\thanks{Authors' addresses: H. Jung and K. Kim, School of Electrical Engineering and Computer Science, Gwangju Institute of Science and Technology, Gwangju, South Korea
(e-mail:\{rain, kskim\}@gist.ac.kr);
 T. S. Lee, Telecommunications Technology Association, Seongnam, South Korea (nason927@tta.or.kr)
 S. Kim, Agency for Defense Development, Daejeon, South Korea (sikim777@add.re.kr)
 J. Kang, Soundly corp., Seoul, South Korea (jwkang@soundl.ly).}

}


\maketitle
\begin{abstract}
We consider a land mobile satellite communication system using spread spectrum techniques where the uplink is exposed to MT jamming attacks, and the downlink is corrupted by multi-path fading channels.
We proposes an anti-jamming receiver, which exploits inherent low-dimensionality of the received signal model, by formulating a robust principal component analysis (Robust PCA)-based recovery problem.
Simulation results verify that the proposed receiver outperforms the conventional receiver for a reasonable rank of the jamming signal.
\end{abstract}


\IEEEpeerreviewmaketitle

\section{Introduction}
\IEEEPARstart{L}{and} mobile satellite (LMS) communications facilitate a myriad of applications such as Global Navigation Satellite System (GNSS) and commercial broadcasting systems, e.g. DVB-RCS and S-DMB \cite{DEK1}.
Due to the openness of LMS communications, both uplink and downlink channels are easily interrupted by unexpected interferences of surrounding communications as well as intentional jammers as reported in \cite{DEK2}.
Furthermore, the frequency selectivity caused by multi-path scatters at the receivers' side makes it difficult to recover the source data under LMS applications operating at high-frequency bands \cite{DEK3}.

Spread spectrum (SS) techniques are data modulation methods that spread the bandwidth of the signal over the bandwidth actually required. 
Systems adopting SS techniques have been effectively utilized for the suppression of interferences, alleviation of multi-path fading effects, and the resilience against jamming signals \cite{DEK4}. 
Code division multiple access (CDMA) provides multiple access capability and helps increasing system throughput by applying the SS notion. 
CDMA is classified into time-hopping (TH), frequency-hopping (FH), and direct sequence (DS) SS, among which DS-CDMA has been mostly studied in the literature and widely applicable in reality due to its low complexity and implementation cost \cite{BMD-DSCDMA, BSS-ICA-CDMA}. 
The performance of DS-CDMA is limited by jamming signals and interferences since they often exceed the anti-jamming capability of SS techniques.

To mitigate the effect of jamming/interference signals, the most common method is to filter the received signal in space, time, and frequency domains \cite{ST-AES-GNSS ,ST-Sen-GNSS ,ANF-ISJ-GPS, TF-AES-GNSS, STFT-IET-GPS}.
Space-time adaptive processing can mitigate wideband and narrowband jamming, but it requires additional antennas \cite{ST-AES-GNSS,ST-Sen-GNSS}. Time-frequency filtering can alleviate narrowband and continuous-wave jamming, it however requires some prior information about jamming signals \cite{ANF-ISJ-GPS, TF-AES-GNSS, STFT-IET-GPS}. 
A main weakness of aforementioned filtering methods is an extreme degradation of the anti-jamming performance when the jamming signals are coming from the same direction with the source signals, and subsequently high jamming-to-signal ratio (JSR). 

In this context, blind source separation (BSS) using independent component analysis (ICA) was proposed to relax requirements \cite{DEK10}. BSS with ICA separates multiple source signals by analyzing the statistical independences using higher order statistics with the assumption that signals from different sources are statistically independent \cite{DEK12}. BSS-ICA provides a wide applicability, including blind multiuser detection, which is to recover the source bit sequence from a received mixture without any knowledge of the user spreading code \cite{BMD-DSCDMA}, and jamming suppression in CDMA communications \cite{BSS-ICA-CDMA,DEK10,DEK11}.
One limitation of BSS-ICA is that it requires a number of observations equal to or larger than a number of sources that we want to separate. Additionally, the anti-jamming capability degrades when the jamming signals are varying in both the time and frequency domain, and the source is already corrupted by jamming in the uplink channel.

In this work, we investigate the anti-jamming behavior of CDMA-based LMS communications, where the satellite acts as a simple amplify-and-forward (AnF) relay.
We consider the uplink jamming scenario, which is frequently used in electronic warfare because it is efficient to impair all receivers critically at once \cite{DEK2}.
The uplink jammer is assumed as a multi-tone (MT) jamming with frequency hopping (FH), which is one of the principal categories of intelligent jamming strategies \cite{MTJ-TWCOM}. We observe that the jamming signals actually rely on a few number of jamming frequencies and a number of hopping occurrences. 
With these observations, the matrix representation of the jamming signal can be modeled as a low-rank matrix having low-dimensionality. Low-rank jamming/interference can be easily found in many emerging applications, including communication and network systems \cite{LRI-TCOM,LRI-TWCOM}. Based on the scenario we discussed, descriptions of the signals and the system, including jamming signals are detailed in Subsection $\mathrm{II}$. $A$.
We also remark that the number of active users is often much less than the multi-user capacity of systems for many applications, including CDMA \cite{SUA-TCOM,RDMUD-TIT,SSP-5G-IAC}. This low activity thus implies sparse DS-CDMA signals having low-dimensionality property. 

The present paper fruitfully exploits low-dimensionality attributes to recover the source signal from the received signal where an MT-FH jamming was interfering in the uplink channel. The approach we propose in this paper is to model the DS-CDMA signal and the jamming signal into matrix representations and to formulate the recovery problem into a matrix decomposition problem. To decompose the received signal by utilizing their low-dimensionality, we suggest an anti-jamming DS-CDMA receiver, including robust principal component analysis (Robust PCA) in addition to ICA based receiver. Robust PCA recovers a matrix $\mathbf{L}$ from highly corrupted measurements $\mathbf{Q=L+R}$, where $\mathbf{L}$ and $\mathbf{R}$ are low-rank and sparse matrices, respectively \cite{DEK7}.
In contrast to Gaussian noise in classical PCA, the entries in a sparse matrix $\mathbf{R}$ can have larger magnitudes which are unknown.
Extensive simulations show that Robust PCA performs better than ICA only under the assumptions of the low-rank jamming signal and the sparsity of a transmitted DS-CDMA signal when a number of users are less than the length of a spreading code. Even in the other cases, the proposed receiver guarantees a comparable anti-jamming performance to ICA only. 

This paper is organized as follows.
Section $\rm{II}$ formulates the system model, uplink scenario, and downlink scenario separately.
Section $\rm{III}$ suggests a recovery problem using matrix decompositions, Robust PCA and ICA, with algorithms to solve the optimization problems.
Section $\rm{IV}$ presents numerical simulation results to justify the anti-jamming ability of the proposed receiver structure.
Finally, section $\rm{V}$ summarizes the paper.

\section{System Model}
The system model considered in this paper consists of a transmitter, a land-based jammer, a satellite, and a receiver.
We divide the system model into two subsections: uplink scenario and downlink scenario.
In the uplink scenario, the transmitted signal and jamming signal models are provided in matrix forms.
In the downlink scenario, the LMS channel is formulated as a circulant matrix, and the received signal model is given.
In what follows, the system model is explained based on the block diagram of proposed anti-jamming CDMA structure depicted in Fig. 1.

%

\begin{figure}[t]
	\centering
	\includegraphics[width=8.8cm]{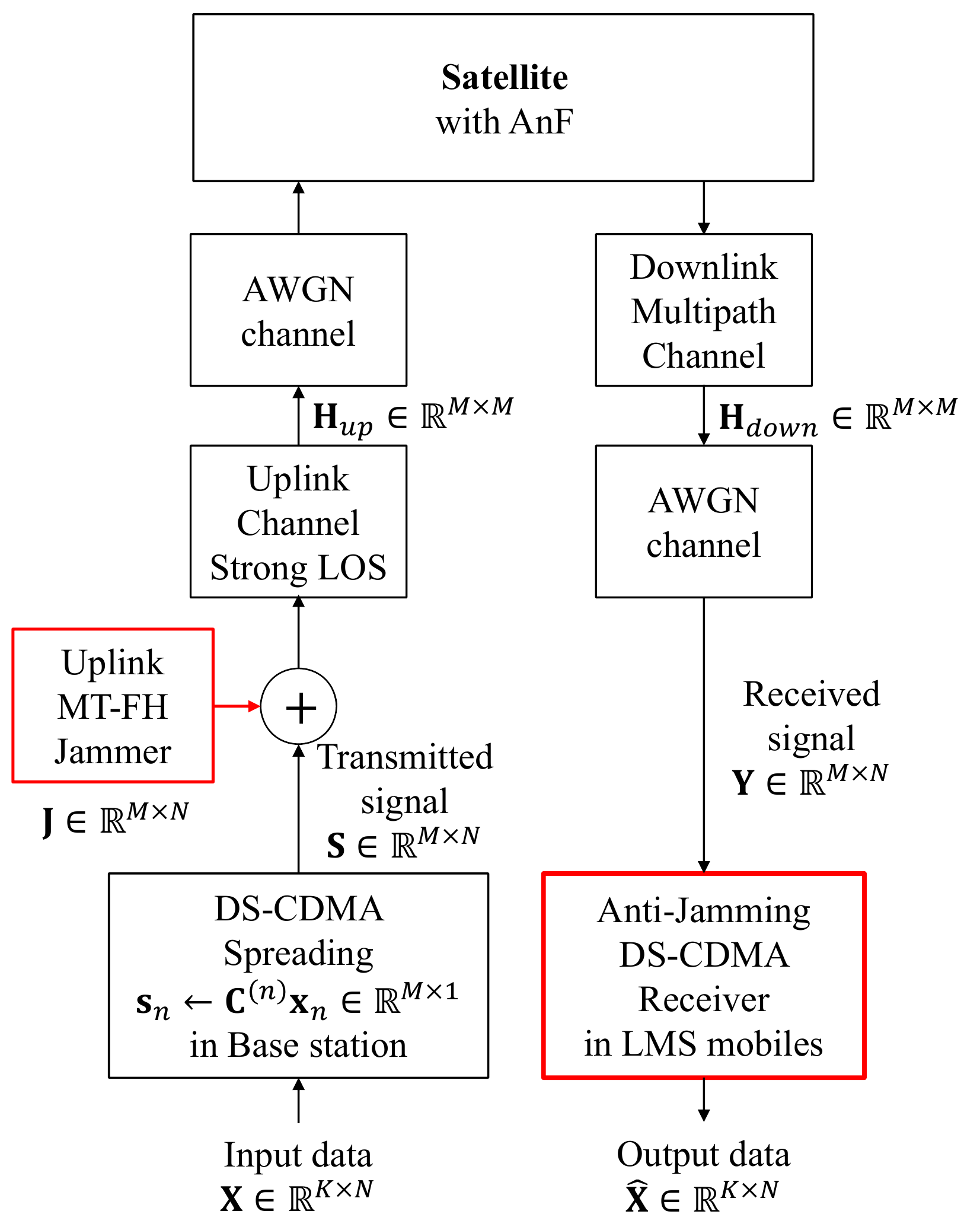}
	\caption{Block diagram of the LMS communication systems with the anti-jamming DS-CDMA receiver}
\end{figure}

\subsection{Uplink Scenario}
On the satellite uplink, for synchronous CDMA transmissions by $K$ multi-users at the base station, the input data ${\mathbf{X}} \in {\mathbb{R}^{K \times N}}$ is given in a matrix form where $K$ users have $N$ bits. 
The input data $\mathbf{X}$ can be divided into $N$ column vectors as follows:
\begin{align}
{\bf{X}} = [{{\bf{x}}_1},\, \cdots,\, {{\bf{x}}_n},\, \cdots,\, {{\bf{x}}_N}] \in {\mathbb{R}^{K \times N}},
\end{align}
where ${\mathbf{x}}_{n}=\left[x_{1,n},\, \cdots,\, x_{k,n},\, \cdots,\, x_{K,n} \right] $ is a column vector that is a collection of $n^{th}$ bits of $K$ users, and $x_{k,n}$ is the $n^{th}$ bit of the $k^{th}$ user.
The transmitted DS-CDMA signal $s(t)$ is represented as \cite{BMD-DSCDMA}:
\begin{align}\label{eq:s_t}
s(t)=\sum_{k=1}^{K}{\sum_{n=1}^{N}{\sum_{m=1}^{M}{x_{k,n}c_k(t-nT_b-mT_c
)}}},
\end{align}
where $c_k(\cdot)$ is the $k^{th}$ user spreading code, $T_c$ is chip duration, $T_b=M T_c$ is the bit duration, and $M$ is the length of the spreading code. Sampling by $T_c$, the encoding of DS-CDMA \eqref{eq:s_t} can be formulated into a matrix representation using the $n^{th}$ spreading code matrix ${{\bf{C}}^{(n)}} \in {\mathbb{R}^{M \times K}}$.
The $N$ numbers of spreading code matrices are generated at every bit index $n=1,\dots,N$, and each spreading code matrix randomly chooses $K$ column vectors from Walsh code $\mathbf{W} \in {\mathbb{R}^{M \times M}}$. Walsh code is adopted out of Gold code, maximal length code, and Walsh code due to its orthogonality and simplicity.

The transmitted signal matrix ${\bf{S}} \in {\mathbb{R}^{M \times N}}$, which is a collection of samples $s_{m,n}=s(mT_c+nT_b)$ is the output of spreading block and is formulated as:
\begin{align}
{\bf{S}} = [{{\bf{s}}_1},\, \cdots,\, {{\bf{s}}_n},\, \cdots,\, {{\bf{s}}_N}] \in {\mathbb{R}^{M \times N}},
\end{align}
where the $n^{th}$ column vector of ${\bf{S}}$ is generated by:
\begin{align}
{\bf{s}}_{n} = {{\bf{C}}^{(n)}}{\bf{x}}_{n} \in {\mathbb{R}^{M \times 1}}.
\end{align}

The transmitted signal ${\bf{S}}$ is jammed by uplink jamming signals.
Typically, jamming signals are characterized by frequency parameters such as jamming frequency bandwidth in partial band jamming and jamming frequencies in MTs jamming. 
In our system model, the MT-FH jamming signal is given as:
\begin{align}
j(t)=\sqrt{\frac{P_J}{Mp}} \sum_{m=1}^{M}{\delta_m(t) \mathrm{exp}[i2\pi f_mt+\phi_m]},
\end{align}
where $P_J$ is the power of the jamming signal, the quantity $\delta_m(t)$ is equal to 1 when the $m^{th}$ frequency is jammed at time $t$ with a probability $p$, $f_m$ is the $m^{th}$ frequency, $\phi_m$ is the phase of the $m^{th}$ tone jammer. It is noted that the power of the jamming signal is divided by $Mp$ for the normalization.

The Fourier transform of the jamming signal $j(t)$, during $n^{th}$ bit duration $[nT_b,nT_b+(m-1)T_c]$, can be represented in a vector  $\mathbf{j}_n^{'}\in \mathbb{R}^{M\times1}$. The $m^{th}$ frequency element of $\mathbf{j}_n^{'}$ is formed as:
\begin{align}
{j}_{m,n}^{'} = \delta_m(n) Z \sqrt{\frac{P_J}{Mp}},
\end{align}
where $\delta_m(n)$ is 1 when the $m^{th}$ frequency is jammed at $n^{th}$ bit duration with a probability $p$, and $Z$ is a random variable that is distributed normally, i.e., $Z\sim \mathcal{N}(0,1)$.

Using the function $\delta_m(n)$, various types of jamming signals, including narrowband, MT, and wideband jamming, can be generated by adjusting non-zero frequency components. The time domain representation of $\mathbf{j}_n^{'}$ is obtained by inverse Fourier transformation, i.e., $\mathbf{j}_n=\mathcal{F}^{-1}\{\mathbf{j}_n^{'}\}$. 
The jamming signal ${\bf{J}} \in {\mathbb{R}^{M \times N}}$ for entire bit durations is given as:
\begin{align}
{\bf{J}} = [{{\bf{j}}_1},\, \cdots,\, {{\bf{j}}_n},\, \cdots,\, {{\bf{j}}_N}] \in {\mathbb{R}^{M \times N}}.
\end{align}

In the case of a typical MT jammer without FH, column vectors of the jamming signal ${\mathbf{J}}$ are same during the whole set of bit durations, which is represented as ${\bf{j}}_{1}={\bf{j}}_{2}=...={\bf{j}}_{N}$. 
In other words, the jammer attacks the same frequency components of all column vectors that is ${\bf{j}}_{1}^{'}={\bf{j}}_{n}^{'}\forall n=1,\dots,N$, and thus, we obtain the jamming signal ${\bf{J}}$, which is a rank-1 matrix.
In addition, we consider an MT-FH jamming signal that the jammer hops the jamming frequency components several times. Consequently, the jammer also changes jamming vectors  ${\bf{j}}_{n}$ depending on their frequency vectors ${\bf{j}}_{n}^{'}$.
If the number of hops increases in the MT-FH jamming signal, the rank of the jamming signal also increases.
For instance, if the jammer hops four times, then it generates four jamming frequency vectors randomly and the jammer transmits the inverse Fourier transform of each jamming frequency vector until the jammer hops their jamming frequencies. Finally, the jamming signal ${\bf{J}}$ has four distinct parts and can be calculated as a rank-4 matrix. Let rank-$r$ denote the rank of the jamming signal and $r$ represent the number of hopping events.
Signal-to-jamming ratio (SJR) is defined as follows:
\begin{align}
\textrm{SJR\ \ [dB]} = 20 \log \frac{{\lVert\mathbf{S}\lVert}_F}{{\lVert\mathbf{J}\lVert}_F}\ \ \textrm{[dB]},
\end{align}
where ${\lVert\mathbf{S}\lVert}_F=\sqrt{\sum_{m=1}^{M}{\sum_{n=1}^{N}{\lvert s_{m,n}\lvert}^2}}$ is the Frobenious norm of a matrix $\mathbf{S}$, which represents the signal energy.

The received signal in the satellite, which is jammed by the jamming signal ${\bf{J}} \in {\mathbb{R}^{M \times N}}$, can be expressed as ${\bf{H}}_{up} * ( {{\bf{S}}} + {{\bf{J}}} ) + {\bf{V}}_{1} \in {\mathbb{R}^{M \times N}}$, where ${\bf{H}}_{up}$ defines the uplink channel assuming that there always exists a highly strong line-of-sight (LOS) path due to the aid of directional antennas pointing to the satellite \cite{DEK2}.
${\bf{V}}_1$ is a simple additive white Gaussian noise (AWGN). We now consider the satellite as a simple AnF relay which amplifies signals by an amplifying gain $G_{AnF}$ and transmits the outcome to the receiver.

\begin{figure}[t]
	\centering
	\includegraphics[width = 8.8cm]{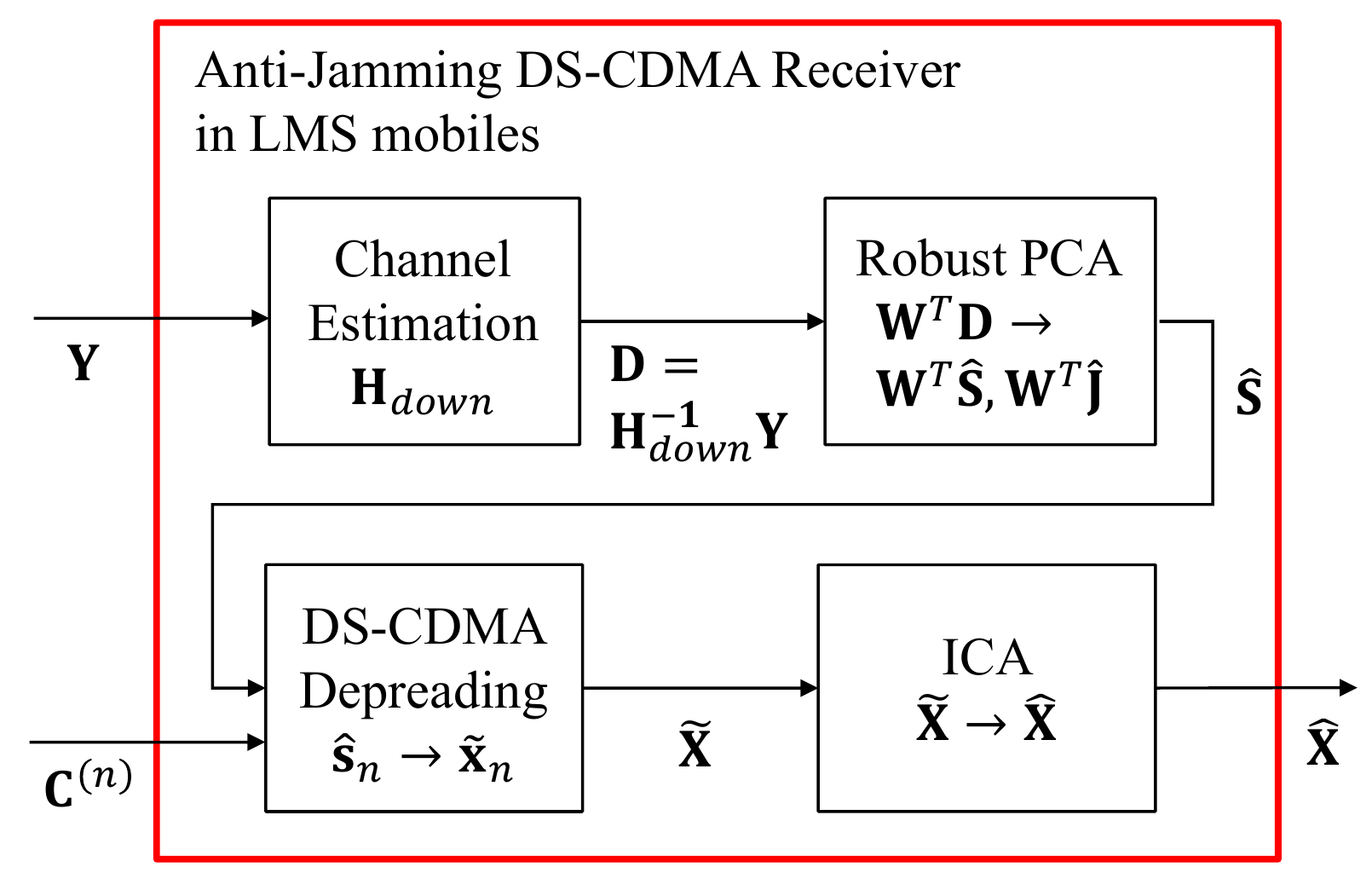}
	\caption{Details of the anti-jamming DS-CDMA receiver blocks from Fig. 1.}
\end{figure}

\subsection{Downlink Scenario}
We assume that the corresponding downlink receiver must be designed with consideration of the LMS characteristics due to the mobility of receivers.
The conventional LMS literature states that such a satellite downlink is represented as a frequency-selective channel consisting of a LOS path and 2 to 4 clustered diffuse paths with high path-loss \cite{DEK3}.
When we express the frequency-selective channel in a discrete form, the channel impulse responses are divided into three components: a direct path, near echoes and far echoes.
We mathematically model the downlink frequency-selective channel using a circulant matrix \cite{DEK13} as given below:
\begin{align}
{\bf{H}}_{down} = {\rm{CM}}[{{h_0}},\, {{h_1}},\, \cdots,\, {{h_l}},\, \cdots,\, {{h_{L - 1}}}
] \in {\mathbb{R}^{M \times M}}.
\end{align}
Let ${\bf{h}} = [{h_0},\, \cdots,\, {h_l},\, \cdots,\, {h_{L - 1}}]$ be the equivalent discrete time channel impulse response, ${\rm{CM}[\cdot]}$ indicates the circulant matrix that begins with the index ${\bf{h}}$, and $L$ denotes the number of paths of the downlink channel.
Discrete channel impulse response ${h_l}$ is a complex Gaussian random variable representing fading channel environments $(l=0,...,L-1)$.
Finally, the received DS-CDMA signal is modeled as:
\begin{align}\label{r_t}
\medmath{{\bf{Y}} = {\bf{H}}_{down} * \{G_{AnF} * {\bf{H}}_{up} * ({\bf{S}} + {\bf{J}})\} + {\bf{V}} \in {\mathbb{R}^{M \times N}},}
\end{align}
where ${\bf{V}}$ denotes the sum of both uplink and downlink AWGN channels whose elements are i.i.d. 
Signal-to-noise ratio (SNR) is defined as a ratio between the average powers of the signal and the AWGN noise as follows:
\begin{align}
	\textrm{SNR\ \ [dB]} = 20 \log \frac{{\lVert\mathbf{S}\lVert}_F}{{\lVert\mathbf{V}\lVert}_F}\ \ \textrm{[dB]}.
\end{align}

Fig. 2 details the proposed recovery block which comprises of four specific blocks: Channel Estimation, Robust PCA, Despreading, and ICA.
This paper deals with practical and diverse downlink channel models representing urban with/without a LOS path, and rural with/without a LOS path environments, which are specified in \cite{DEK3}.
We assume that the receiver has a perfect knowledge of the spreading code ${\bf{W}} \in {\mathbb{R}^{M \times M}}$. In addition, perfect channel estimation of the downlink channel matrix $\widehat{\mathbf{H}}_{down}={\bf{H}}_{down} \in {\mathbb{R}^{M \times M}}$ is also considered.


\section{Recovery Problem \& with Matrix Decomposition}
In this section, we describe a recovery problem of the received signal of \eqref{r_t}, and formulate the recovery problem as a matrix decomposition problem. Our approach then is to decompose the received signal $\mathbf{Y}$ into the transmitted signal $\mathbf{S}$ and the jamming signal $\mathbf{J}$ by utilizing their inherent low-dimensionality features.  

\subsection{Low-Dimensionality Properties \& Channel Estimation}
We demonstrate low-dimensionality properties of the transmitted DS-CDMA signal matrix and the uplink jamming signal matrix that they can be often represented as a sparse matrix and a low-rank matrix in many emerging applications \cite{SUA-TCOM, SSP-5G-IAC, RDMUD-TIT, LRI-TCOM, LRI-TWCOM}.

First, the transmitted DS-CDMA signal $\mathbf{S}\in\mathbb{R}^{M\times N}$ has low-dimensionality, since the number of active users in CDMA systems is often much lower than the spreading gain ($K\ll M$) \cite{SUA-TCOM, RDMUD-TIT}. This low activity is observed in a wide range of applications. In typical tactical communications, active users are usually very small because the spreading gain of military systems focuses mainly on the anti-jamming capability. Due to emerging 5G and IoT technologies, numerous devices are inactive most of the time but occasionally communicate for minor updates \cite{SSP-5G-IAC}. 
If the receiver has a priori information of spreading codes as the transmitter, each column in matrix ${{\bf{W}}^{T}\bf{S}} \in {\mathbb{R}^{M \times N}}$ has only $K$ numbers of non-zero components due to column vectors of Walsh code that are generated independently.
Therefore, the low-dimensionality of the DS-CDMA signal matrix can be represented by the sparse matrix ${{\bf{W}}^{T}\bf{S}} \in {\mathbb{R}^{M \times N}}$.

Second, we also remark that low-rank jamming signals are present and studied in \cite{LRI-TCOM, LRI-TWCOM} and references therein. With this observation, the MT-FH uplink jamming signal matrix ${\bf{J}} \in {\mathbb{R}^{M \times N}}$ can be assumed to have the low-dimensionality. 
Aforementioned in Subsection $\rm{II}$. $A$, the typical MT-FH jamming signal matrix ${\bf{J}} \in {\mathbb{R}^{M \times N}}$ is modeled as the low-rank matrix.
The term ``low-rank matrix" refers to a rank of the matrix that is small compared to the largest possible rank.
Moreover, since $rank(AB) \le \min \{ rank(A),rank(B)\}$, a matrix ${{\bf{W}}^{T}\bf{J}} \in {\mathbb{R}^{M \times N}}$ is also a low-rank matrix.

Depending on the above descriptions, our objective is to propose an anti-jamming DS-CDMA recovery structure, which is depicted in Fig. 2, exploiting the low-dimensionality of the transmitted signal and the jamming signal.
We assume that the AnF gain of the satellite compensates the uplink channel, i.e., $G_{AnF}*\mathbf{H}_{up}=\mathbf{I}_M\in \mathbb{R}^{M \times M}$, where $\mathbf{I}_M$ is an identity matrix with size $M$. This assumption is due to the strong LOS path in the uplink channel.
With the assumption of the perfect estimation $\mathbf{H}_{down}$, Robust PCA decomposes  ${{\bf{W}}^{T}\bf{S}}$ and  ${{\bf{W}}^{T}\bf{J}}$ from ${{\bf{W}}^{T}\bf{D}}$, where ${\bf{D}}={\bf{H}}_{down}^{-1} {\bf{Y}}$.
The input and output signals of the Robust PCA block are ${\bf{D}} \in {\mathbb{R}^{M \times N}}$ and ${\widehat{\bf{S}}} \in {\mathbb{R}^{M \times N}}$, respectively.
The Despreading block then despreads ${\widehat{\bf{S}}} \in {\mathbb{R}^{M \times N}}$ with the known spreading code matrices for all bits ${{\bf{C}}^{(n)}} \in {\mathbb{R}^{M \times K}} \forall n=1,\dots,N$.
Finally, ICA reconstructs the original signal ${\widehat{\bf{X}}} \in {\mathbb{R}^{K \times N}}$ from ${\widetilde{\bf{X}}} \in {\mathbb{R}^{K \times N}}$ by using independence inherently contained in the received signal.

In subsections $\rm{III}$. $B$ and $\rm{III}$. $C$, we delineate the functionality of the recovery block regarding matrix decomposition.
To implementing Robust PCA and ICA for our anti-jamming DS-CDMA receiver, we modify iALM and Fast ICA algorithms for the system model of this paper.

\begin{algorithm}[t]
	\DontPrintSemicolon
	\caption{iALM for Robust PCA problem}\label{alg:iALM}
	\KwData{${{\bf{W}}^{{T}}}{\bf{D}} \in {\mathbb{R}^{M \times N}},{\rm{ }}\lambda  = 1/\sqrt M $}
	\KwResult{~~$\mathbf{W}^T\widehat{\bf{J}} \leftarrow \mathbf{L}_k, \mathbf{W}^T\widehat{\mathbf{S}} \leftarrow \mathbf{R}_{k}$}
	\BlankLine
	${\bf{\Lambda }} _{0} \leftarrow {{\bf{W}}^{{T}}}{\bf{D}}/ \max  \left( \lVert{{\bf{W}}^{{T}}}{\bf{D}}\rVert _2 ,\lambda^{-1}\lVert{{\bf{W}}^{{T}}}{\bf{D}}\rVert_\infty\right).$\;
	${\bf{R}}_{0} \leftarrow 0$, ${\mu_0} \leftarrow {1.25}/{{\lVert {\bf{W}}^{{T}}}{\bf{D}\rVert}_2}$, $k \leftarrow 0$.\;

	\While{not converged}{
		\tcp*[l]{Solve $\mathbf{L}_{k+1}=\arg{\min\limits_{\mathbf{L}}{L(\mathbf{L}, \mathbf{R}_{k}, \Lambda_{k}, \mu_{k})}}$}
		$[{\bf{U,P,V}}] \leftarrow {\mathrm{svd}}({{\bf{W}}^{{T}}}{\bf{D}} - {\bf{R}}_{k} + {\mu} _{k}^{ -1}{\bf{\Lambda }} _{k})$.\;
		${\bf{L}}_{k+1} \leftarrow {\bf{U}}\cdot{\mathrm{Th}}[{\bf{P}}:{{\mu} _{k}^{ - 1}}]\cdot{{\bf{V}}^{{T}}}$.\;
		\tcp*[l]{Solve $\mathbf{R}_{k+1}=\arg{\min\limits_{\mathbf{R}}{L(\mathbf{L}_{k+1}, \mathbf{R}, \Lambda_{k}, \mu_{k})}}$}
		${\bf{R}}_{k+1} \leftarrow {\mathrm{Th}}[{{\bf{W}}^{{T}}}{\bf{D}} - {\bf{L}}_{k+1} + {\mu} _{k}^{ - 1}{\bf{\Lambda }} _{k}:\lambda{{\mu} _{k}^{ - 1}}] $.\;
		${\bf{\Lambda }} _{k+1} \leftarrow {\bf{\Lambda }} _{k} + {\mu _k}({{\bf{W}}^{{T}}}{\bf{D}} - {\bf{L}}_{k+1} - {\bf{R}}_{k+1})$.\;
		Update $\mu_{k}$ to $\mu_{k+1}$. \;
		$k \leftarrow k + 1$.\;
	}
\end{algorithm}

\subsection{The iALM Algorithm for Robust PCA}
We now consider a matrix decomposition problem to recover the sparse DS-CDMA signal ${{\bf{W}}^{T}\bf{S}}$ and the low-rank jamming signal ${{\bf{W}}^{T}\bf{J}}$ by solving the following convex optimization problem: 
\begin{equation}\label{eq:RPCA}
\begin{aligned}
&\mathop {\min }\limits_{{{\bf{W}}^{T}}{\bf{J}},\,{{\bf{W}}^{T}}\bf{S}}~~{\left\| {{{\bf{W}}^{{T}}}{\bf{J}}} \right\|_*} + \lambda {\left\| {{{\bf{W}}^{{T}}}{\bf{S}}} \right\|_1},\\
&{\rm{subject~to}}~~{{\bf{W}}^{{T}}}{\bf{D}} = {{\bf{W}}^{{T}}}{\bf{J}} + {{\bf{W}}^{{T}}}{\bf{S}},
\end{aligned}
\end{equation}
where $\lambda$ is a weighting parameter, ${\left\|  \mathbf{A}  \right\|_1}:=\sum_{m,n}\lvert a_{m,n}\rvert$ denotes the $\ell_1$-norm (i.e., the sum of absolute values of all entries of the matrix $\mathbf{A}$), and ${\left\|  \mathbf{A}  \right\|_*}:=\sum_i \sigma_i(\mathbf{A})$ the nuclear norm of the matrix $\mathbf{A}$(i.e., the sum of singular values of $\mathbf{A}$). The optimization problem \eqref{eq:RPCA} simply minimizes a weighted combination of the nuclear norm and $\ell_1$-norm is referred to as Robust PCA \cite{DEK7}. Robust PCA can recover sparse components of the signal matrix even though the matrix are entirely corrupted by a low-rank matrix. The weighting parameter $\lambda$ controls the balance of regularization between the sparsity and the low-rank constraints. 
Based on prior knowledge to the solution, a choice of $\lambda$ often improves performance. For example, if we know that $\mathbf{W}^T\mathbf{S}$ is very sparse, it is possible to recover matrices $\mathbf{W}^T\mathbf{J}$ of larger rank by increasing $\lambda$. However, $\lambda={1}/{\sqrt{M}}$ is recommended for the existence and the uniqueness of the solution in practical problems \cite{DEK7}. We also choose $\lambda=1/{\sqrt{M}}$ in this paper.

In this paper, we have chosen to solve the Robust PCA problem \eqref{eq:RPCA} using an augmented Lagrangian multiplier (ALM) algorithm introduced in \cite{DEK8}. 
ALM has been proved to converge to the exact optimal solution in fewer iterations \cite{DEK9}.
In practical applications, it works stably across a wide range of problem settings with no parameter tuning \cite{DEK7}.
The ALM method operates on the augmented Lagrangian function of the Robust PCA optimization \eqref{eq:RPCA}
\begin{equation}
\begin{aligned}
L({{\bf{W}}^{{T}}}{\bf{J}},{{\bf{W}}^{{T}}}{\bf{S}},\mathbf{\Lambda} ,\mu )
& \buildrel\textstyle.\over=  {\left\| {{{\bf{W}}^{{T}}}{\bf{J}}} \right\|_*} + \mathbf{\lambda} {\left\| {{{\bf{W}}^{{T}}}{\bf{S}}} \right\|_1}\\
&+ \medmath{\left\langle {\mathbf{\Lambda} ,{{\bf{W}}^{{T}}}{\bf{D}} - {{\bf{W}}^{{T}}}({\bf{S}} + {\bf{J)}}} \right\rangle} \\
\label{eq:ALMF}&+ \medmath{\frac{\mu }{2}\left\| {{{\bf{W}}^{{T}}}{\bf{D}} - {{\bf{W}}^{{T}}}({\bf{S}} + {\bf{J)}}} \right\|_F^2,}
\end{aligned}
\end{equation}
where $\left\langle {A,B} \right\rangle  = {\rm{tr}}({A^T}B)$ and $\mu$ is a positive scalar with a Lagrange multiplier matrix $\mathbf{\Lambda} $. A generic ALM algorithm is to solve \eqref{eq:RPCA} by repeatedly solving
\begin{align}\label{eq:eALM}
\medmath{(\mathbf{W}^T\mathbf{J}_k, \mathbf{W}^T\mathbf{S}_k)=\mathop{\arg \min}\limits_{\mathbf{W}^T\mathbf{J}, \mathbf{W}^T\mathbf{S}}L(\mathbf{W}^T\mathbf{J}, \mathbf{W}^T\mathbf{S}, \mathbf{\Lambda}_k, \mu_{k}),}
\end{align} 
and then update the Lagrangian multiplier matrix by
\begin{align}\label{eq:LMup}
\mathbf{\Lambda}_{k+1}=\mathbf{\Lambda}_k+\mu_{k}(\mathbf{W}^T\mathbf{D}-\mathbf{W}^T(\mathbf{S+J})).
\end{align}
For the low-rank and sparse decomposition problem, the solution of a complex optimization of \eqref{eq:eALM} can be obtained by solving very simple calculations sequentially as follows:
\begin{equation}\label{eq:Ssub}
\begin{aligned}
\mathbf{W}^{T}\mathbf{S}_{k+1} &=\mathop{\arg \min}\limits_{\mathbf{W}^{T}\mathbf{S}} L(\mathbf{W}^{T}\mathbf{J}_{k}, \mathbf{W}^{T}\mathbf{S}, \mathbf{\Lambda}_{k}, \mu_{k})\\
&=\medmath{\mathrm{Th} \left[\mathbf{W}^{T}\mathbf{D}-\mathbf{W}^{T}\mathbf{J}+\mu_{k}^{-1}\mathbf{\Lambda}_{k}:\lambda\mu_{k}^{-1} \right]} ,
\end{aligned}
\end{equation}
\begin{equation}\label{eq:Jsub}
\begin{aligned}
\mathbf{W}^{T}\mathbf{J}_{k+1} &=\mathop{\arg \min}\limits_{\mathbf{W}^{T}\mathbf{J}} L(\mathbf{W}^T\mathbf{J}, \mathbf{W}^{T}\mathbf{S}_{k}, \mathbf{\Lambda}_{k}, \mu_{k})\\
&= \mathbf{U} \cdot \mathrm{Th}\left[\mathbf{P}:\mu_{k}^{-1}\right] \cdot \mathbf{V}^{T},
\end{aligned}
\end{equation} 
where $\mathrm{Th}\left[a:\mu \right]=\mathrm{sgn}(a)\max(\lvert a\rvert-\mu,0)$ is the shrinkage operator and extend it to matrices by applying it to each element, and $\mathbf{UP}\mathbf{V}^{T}=\left[\mathbf{W}^{T}\mathbf{D}-\mathbf{W}^{T}\mathbf{S}_{k}-\mu_{k}^{-1}\mathbf{\Lambda}_{k}\right]$ is any singular value decomposition. In \eqref{eq:Jsub}, the rank of $\mathbf{W}^{T}\mathbf{J}_{k+1}$ is minimized by thresholding corresponding singular values. Additionally, In \eqref{eq:Ssub}, reliable sparse components remain after thresholding elements values.

Algorithm \ref{alg:iALM} describes procedures to solve Robust PCA with proper initialization. Algorithm \ref{alg:iALM} is referred to as inexact ALM (iALM) since it inexactly solves \eqref{eq:eALM} by updating \eqref{eq:Ssub} and \eqref{eq:Jsub} iteratively. Finally, the sparse DS-CDMA signal ${{\bf{W}}^{T}\bf{S}}$ and the low-rank jamming signal ${{\bf{W}}^{T}\bf{J}}$ are decomposed by applying iALM. The initialization of $\mathbf{\Lambda}_0$ in the algorithm is selected to make the objective function \eqref{eq:ALMF} reasonably large. The most important implementation detail of the algorithm is  the choice of $\{\mu_{k}\}$. The choice is directly related to the convergence of the algorithm. It is known that Algorithm \ref{alg:iALM} converges to the optimal solution of Robust PCA if $\{\mu_{k}\}$ is nondecreasing and $\sum_{k=1}^{+\infty}\mu_{k}^{-1}=+\infty$ \cite{DEK9}. We have chosen  $\mu_{0}=1.25/{{\lVert {\bf{W}}^{{T}}}{\bf{D}\rVert}_2}$ and $\mu_{k+1}=\min(1.5\mu_{k}, 10^7 \mu_{0} ),$ where ${{\lVert \mathbf{A}\rVert}_2}=\max_{i}{\sigma_{i}(\mathbf{A})}$ is a 2-norm of a matrix $\mathbf{A}$ i.e., the largest singular value of the matrix $\mathbf{A}$.

After recovering the transmitted signal $\widehat{\mathbf{S}}=\mathbf{W}\mathbf{R}_{k}$ from the received signal by using Robust PCA approach. The MT-FH uplink jamming signal can be removed effectively.
Then, ${\widetilde{\bf{X}}} \in {\mathbb{R}^{K \times N}}$ is obtained by despreading $\widehat{\mathbf{S}}$ with the spreading code matrix $\mathbf{C}^{(n)}$ in Fig. 2.

\begin{algorithm}[t]
\DontPrintSemicolon
	\KwData{${\bf{\widetilde X}}: = ({{{\widetilde x}_{i,j}}}) \in {\mathbb{R}^{K \times N}}$\;
	~~~~~~~~(where $i = 1, \ldots ,K$, and $j = 1, \ldots ,N$)}
	\KwResult{${\widehat{\bf{X}}} \leftarrow {{\mathbf{\omega}}^{{T}}}{\widetilde{\bf{X}}}$} 
	${{{\widetilde x}_{i,j}}}\leftarrow{{{\widetilde x}_{i,j}}}-\frac{1}{N}\sum\limits_{j = 1}^N {{{\widetilde x}_{i,j}}}$ \tcp*[r]{Centering the data}
	$[{\bf{Q}},{\bf{\Gamma}}] \leftarrow {\rm{eig}}({\mathop{\rm cov}} ({\bf{\widetilde X}}))$\;
	${\bf{\widetilde X}} \leftarrow {\bf{Q}}{{\bf{\Gamma}}^{ - 1/2}}{{\bf{Q}}^T}{\bf{\widetilde X}} $ \tcp*[r]{Whitening the data}
	To find initial (random) weight vector ${{\bf{\omega}}_{0}}$; $k = 0$\;
	\While{not converged}{
		${{\bf{\omega}}_{k}} \leftarrow E\{ {\widetilde{\bf{X}}}g{({{\bf{\omega}}_{k}^{{T}}}{\widetilde{\bf{X}}})^{{T}}}\}  - E\{ g'({{\bf{\omega}}_{k}^{{T}}}{\widetilde{\bf{X}}})\} {\bf{\omega}}_{k}$\;
		\tcp*[h]{where $E\{  \cdot \}$ means averaging over\\ all column vectors of matrix ${\bf{\widetilde X}})$}
		${\bf{\omega}}_{k+1} \leftarrow {{\bf{\omega}}}_{k}/\left\| {{{\bf{\omega}}_{k}}} \right\|$\;
		$k \leftarrow k + 1$\;
	}
	\caption{Fast ICA for ICA problem}\label{alg:fICA}
\end{algorithm}

\subsection{Fast ICA Algorithm for ICA}
The next step of our AJ receiver structure is the ICA block which reconstructs the final estimate of the input data ${\widehat{\bf{X}}} \in {\mathbb{R}^{K \times N}}$ from a mixed observation ${\widetilde{\bf{X}}} \in {\mathbb{R}^{K \times N}}$.
BSS using ICA here cannot only detect multi-user signals, but also suppress multi-user interferences, inter-symbol interferences, and intentional jamming signals in CDMA systems \cite{BSS-ICA-CDMA,DEK10}. 
Authors of \cite{DEK10} evaluate the anti-jamming performance of the receiver and show via numerical results that 5dB SJR gains in terms of bit-error-ratio (BER) of $10^{-3}$ under AWGN channel when signal-to-noise-ratio (SNR) is fixed to 20dB.
In our scenario, ICA reconstructs the original signal ${\widehat{\bf{X}}} \in {\mathbb{R}^{K \times N}}$ from ${\widetilde{\bf{X}}} \in {\mathbb{R}^{K \times N}}$, which is also shown in Fig. 2.

To extract independent components from the mixture matrix, we adapt the Fast ICA algorithm \cite{DEK12} which is based on a fixed-point iteration.
For computational simplicity and fast convergence, many studies (also in \cite{DEK10,BMD-DSCDMA}) consider Fast ICA, which is the most popular ICA algorithm thus far.
The Fast ICA algorithm used to restructure ${\widehat{\bf{X}}} \in {\mathbb{R}^{K \times N}}$ is described in Algorithm \ref{alg:fICA}, where $ g(a) = \tanh (a)$ and $g'(a) = 1 - \tanh^{2} (a)$.
The notation ${\mathop{\rm cov}} ({\bf{A}})$ symbolizes the covariance matrix of ${\bf{A}}$. A procedure $[{\bf{Q}},{\bf{\Gamma}}]={\rm{eig}}({\bf{A}})$ performs eigendecomposition of a matrix $\mathbf{A}=\mathbf{Q\Gamma}\mathbf{Q}^{-1}$, where $\mathbf{Q}$ is the square matrix whose columns vectors are eigenvectors of $\mathbf{A}$, and $\mathbf{\Gamma}$ is the diagonal matrix whose diagonal elements are the corresponding eigenvalues.
Fast ICA effectively separates the input data ${\widehat{\bf{X}}}$ by finding an inverse transformation ${{\bf{\omega}}^{T}}{\widetilde{\bf{X}}}$ that maximizes the statistical independence.

In the next section, we perform extensive simulations to verify the anti-jamming ability of the proposed receiver.


\section{Simulation Results and Discussions}
The anti-jamming DS-CDMA receivers using matrix decomposition methods such as Robust PCA and ICA are assessed through simulations for the following two receiver types:

\begin{itemize}
	\item Receiver-Type1 : The conventional anti-jamming DS-CDMA receiver using ICA without Robust PCA,
	\item Receiver-Type2 : The proposed anti-jamming DS-CDMA receiver using both, ICA and Robust PCA approaches.
\end{itemize}

\begin{figure}[t]
	\centering
	\includegraphics[width = 8.6cm]{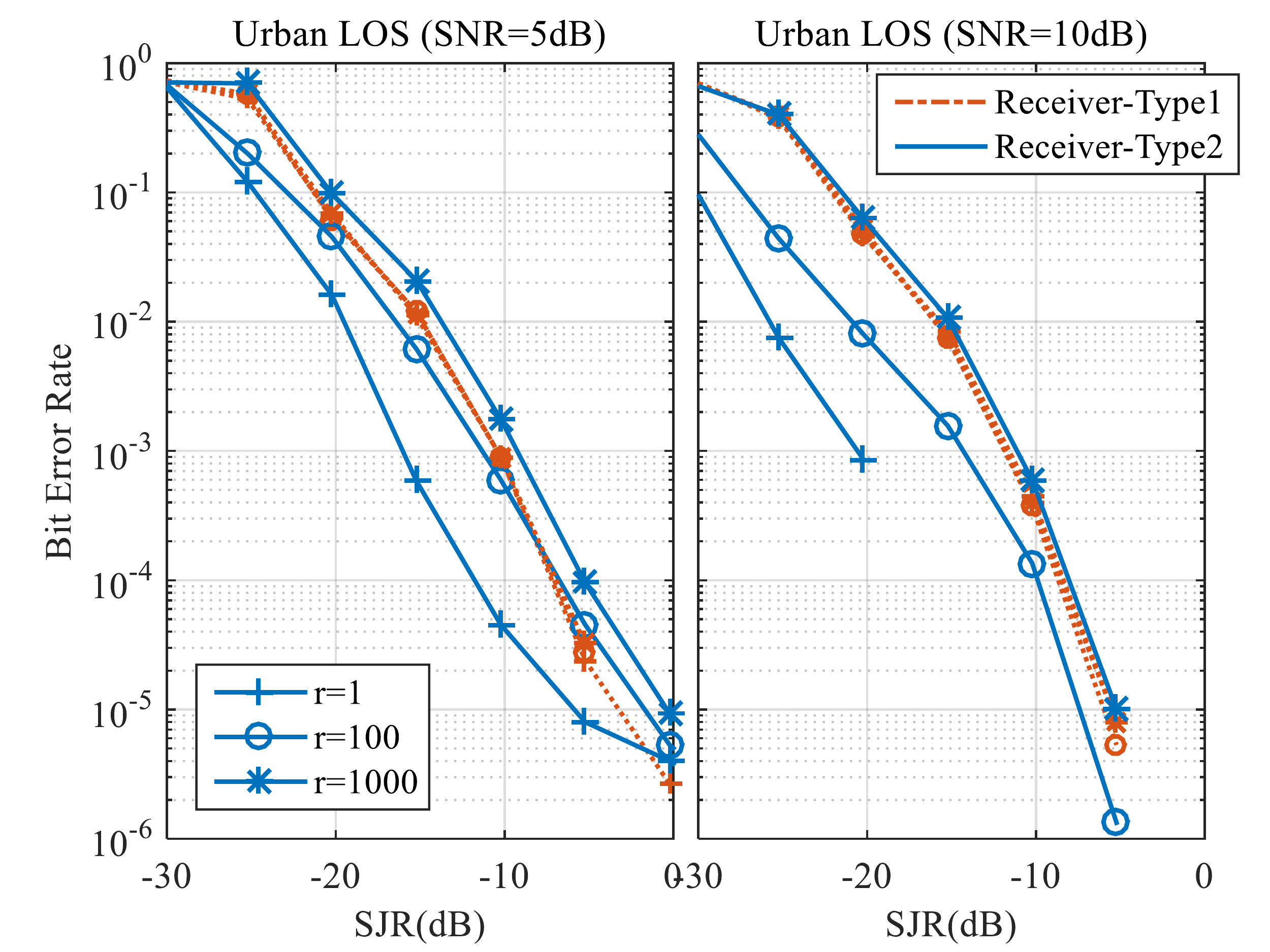}
	\caption{BER versus SJR with SNR fixed to 5dB and 10dB under urban environments (LOS).}
\end{figure}

\begin{figure}[t]
	\centering
	\includegraphics[width = 8.6cm]{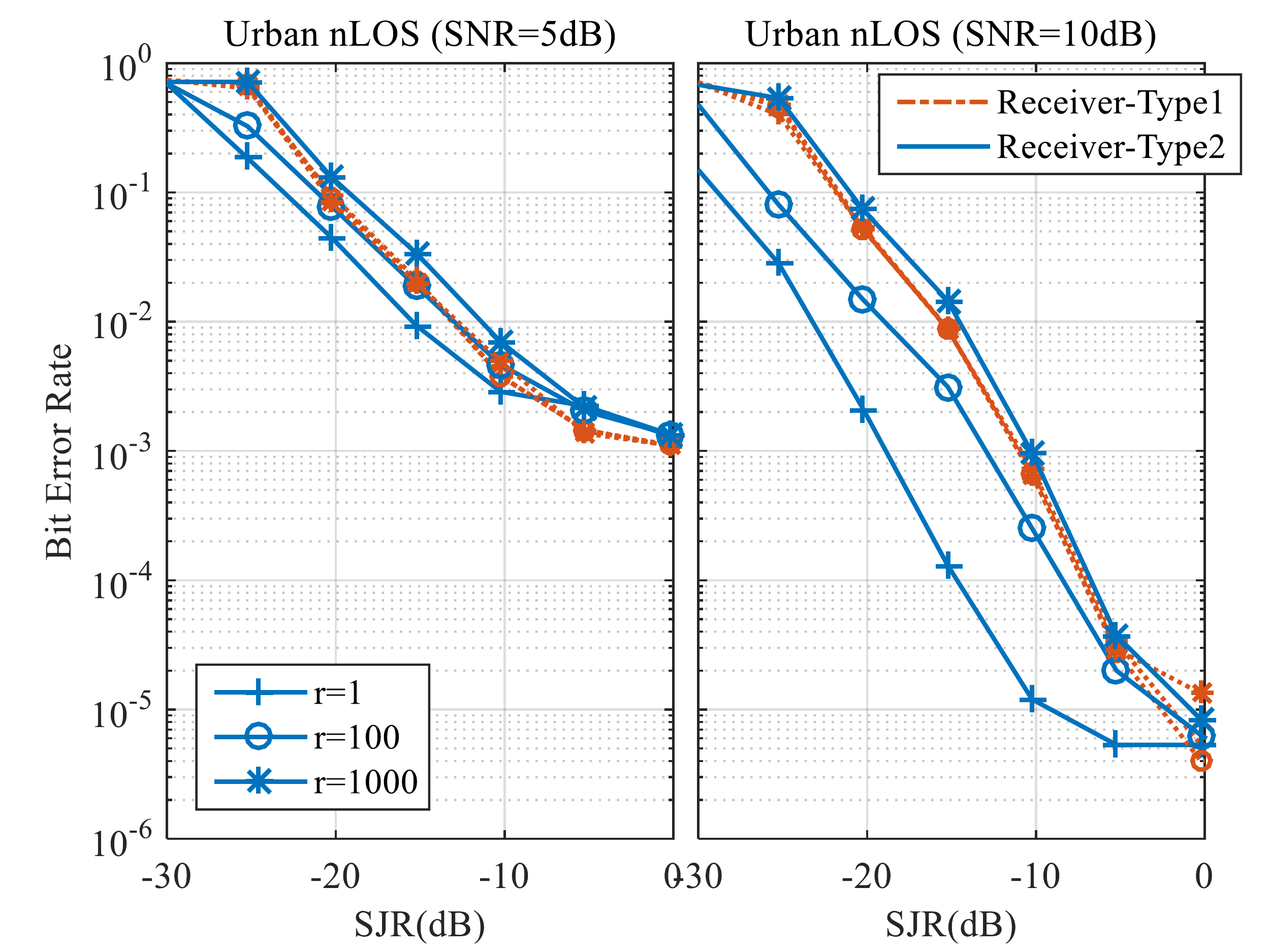}
	\caption{BER versus SJR with SNR fixed to 5dB and 10dB under urban environments (nLOS).}
\end{figure}

\begin{figure}[t]
	\centering
	\includegraphics[width = 8.6cm]{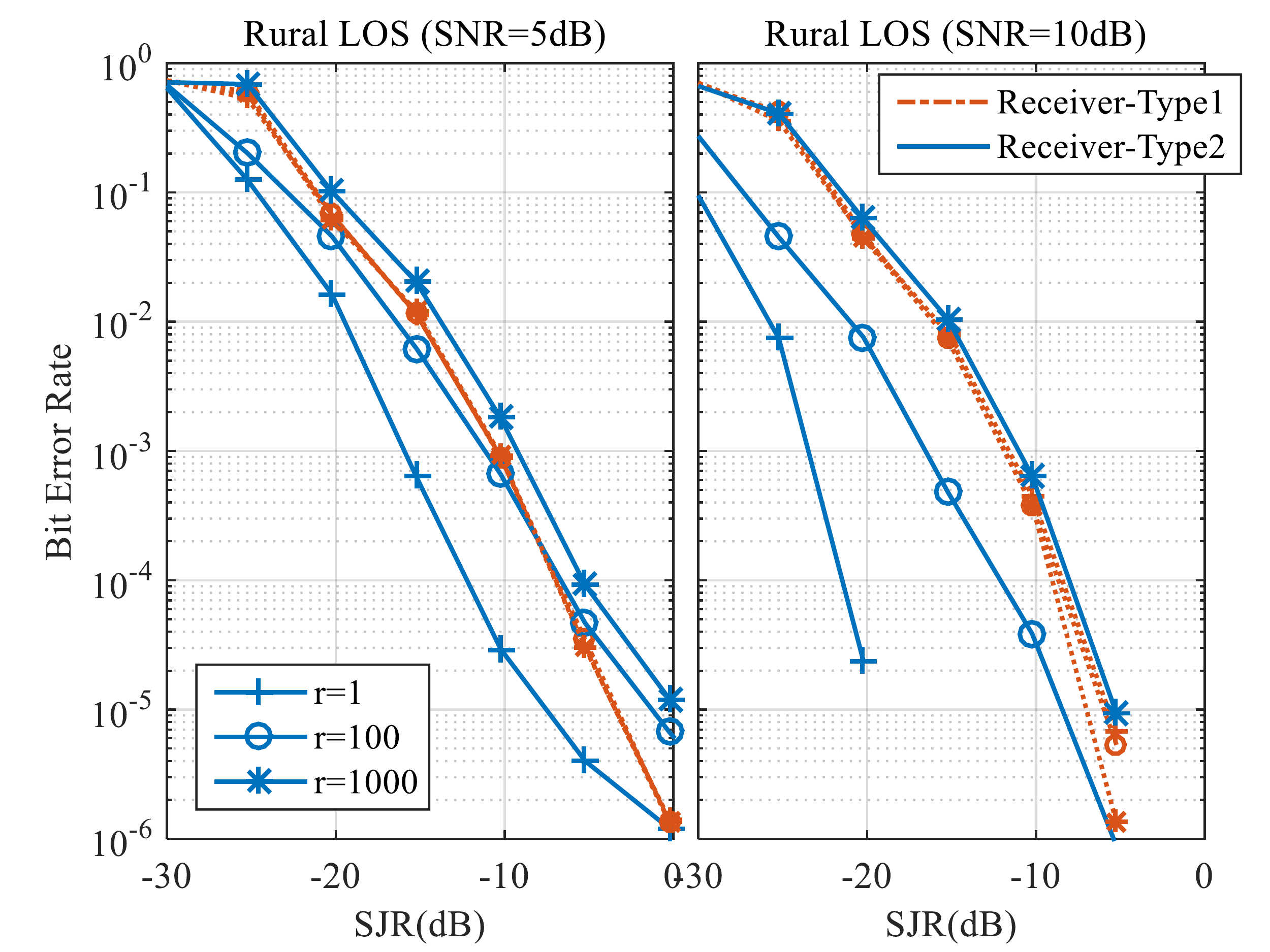}
	\caption{BER versus SJR with SNR fixed to 5dB and 10dB under rural environments (LOS).}
\end{figure}

\begin{figure}[t]
	\centering
	\includegraphics[width = 8.6cm]{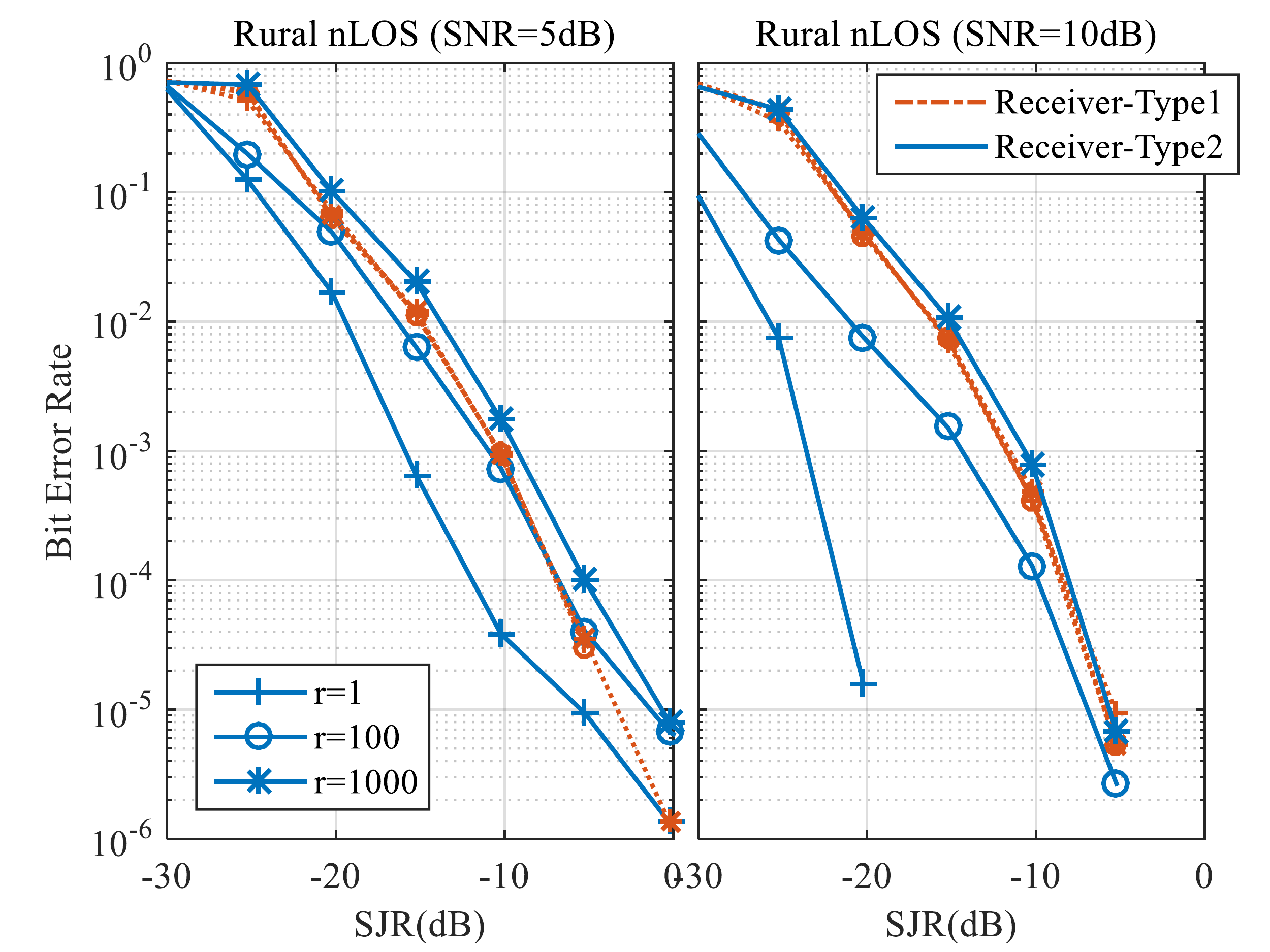}
	\caption{BER versus SJR with SNR fixed to 5dB and 10dB under rural environments (nLOS).}
\end{figure}

The DS-CDMA transmitted signals are generated by following system parameters: $K=30$ users, $N=1000$ bits, and $M=1024$ spreading code length of the Walsh code. The system transmits $M$ chips within each bit duration bearing the information of $K$ users.
We consider various types of the downlink channels such as urban environments with a LOS path and without a LOS path (nLOS), and rural environments with LOS/nLOS. It is known that the downlink channel in LMS communications is a frequency selective channel due to its multi-path propagation consisting of a direct path, near echoes, and far echoes. Parameter sets including a number of taps, delays, and channel gains are set to the measurement data of the LMS International Telecommunication Union (ITU) model \cite{DEK3}.
In the MT-FH uplink jamming scenario, the probability $p$ that the $m^{th}$ frequency is jammed at the $n^{th}$ bit duration is set to be 0.1 i.e., $p=0.1$. Simulations also consider a range of rank-$r$ MT-FH jamming signals such as 1, 100, 200, 500, and at most 1000. The rank of the MT-FH jamming represents the number of hopping events. The case of $r=1$ is a typical MT jamming without hopping, and $r=1000$ is an MT-FH jamming with hopping every bit duration. 
We run 1000 Monte Carlo simulations to observe a reliable BER level of $10^{-5}$ with $\textrm{SJR} =\left[-30,0\right] \textrm{dB}$ and $\textrm{SNR}=5 \textrm{ and } 10\, \textrm{dB}$, as used in \cite{BSS-ICA-CDMA, DEK10}.
It is worthy noted that a broad-band noise jamming can be more effective than the MT-FH jamming as the SJR is too low. However, in the paper, we focus on the MT-FH jamming in order to discuss the effects of the jamming rank $r$ on the performance of the proposed receiver.

Fig. 3, 4, 5 and 6 show BER performances of the Receiver-Type1 and the Receiver-Type2 versus SJR values with various ranks of the MT-FH jamming signal under four different channel scenarios.
The simulation results in Fig. 3 and 4 consider urban environments with the 5 paths frequency selective downlink channels, while Fig. 5 and 6 present the BER performance in rural environments with 3 paths.
Furthermore, Fig. 3 and 5 consider the presence of a LOS path, whereas Fig. 4 and 6 do not.

\begin{figure}[t]
	\centering
	\includegraphics[width = 8.6cm]{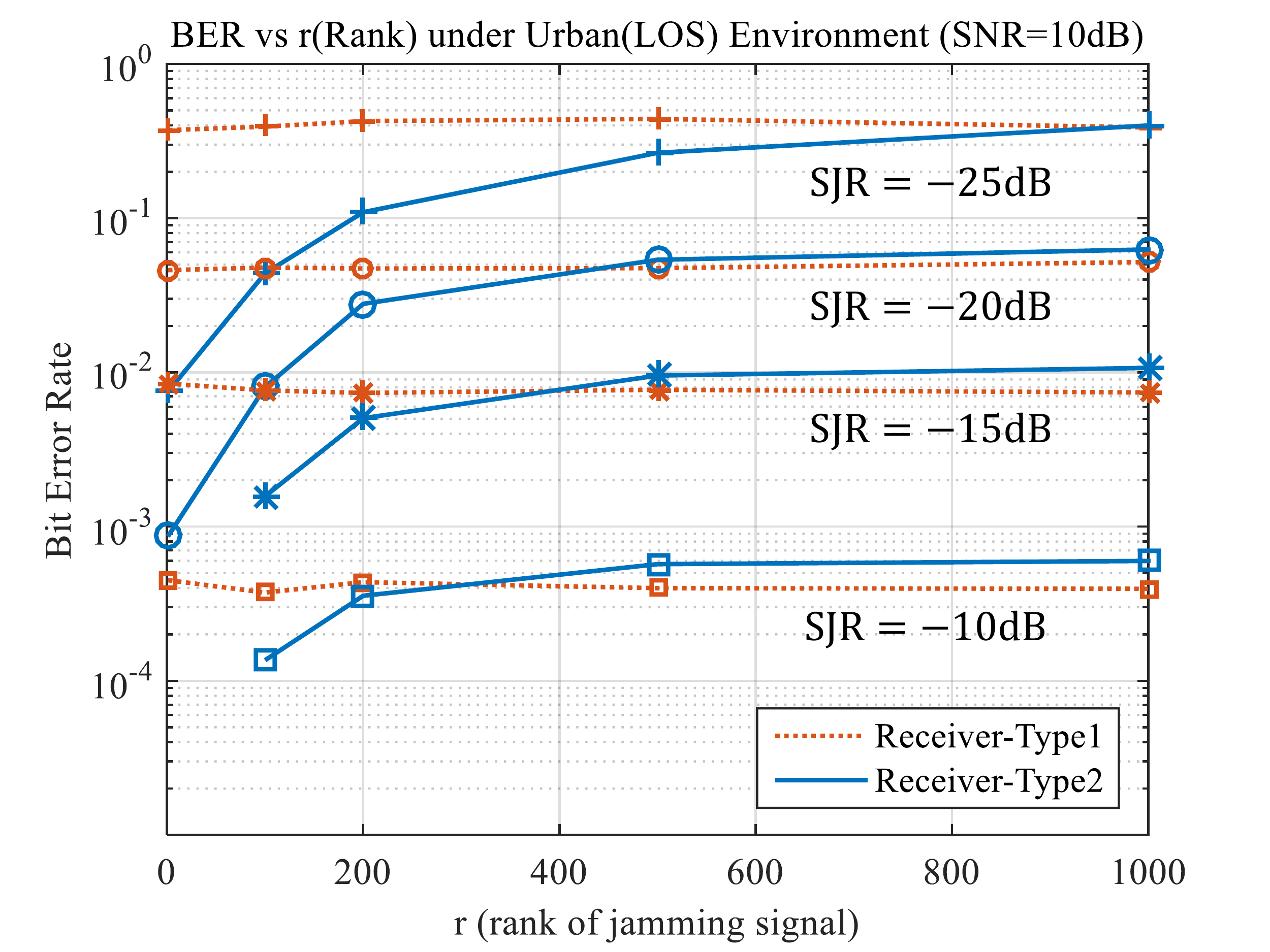}
	\caption{BER versus $r$ (rank of jamming signal) change with SNR fixed to 10dB under urban environments (LOS).}
\end{figure}

\begin{figure}[t]
	\centering
	\includegraphics[width = 8.6cm]{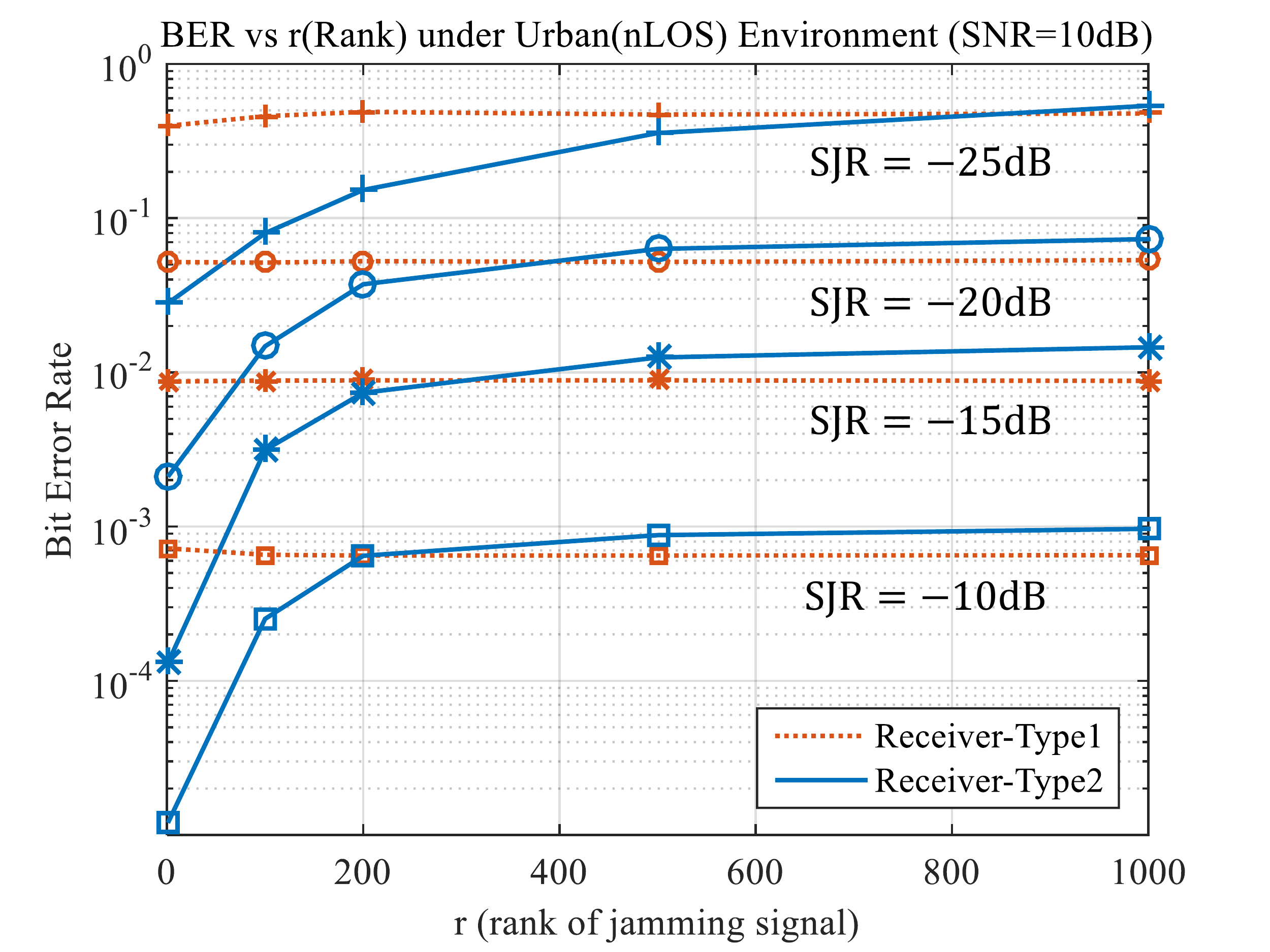}
	\caption{BER versus $r$ (rank of jamming signal) change with SNR fixed to 10dB under urban environments (nLOS).}
\end{figure}

\begin{figure}[t]
	\centering
	\includegraphics[width = 8.6cm]{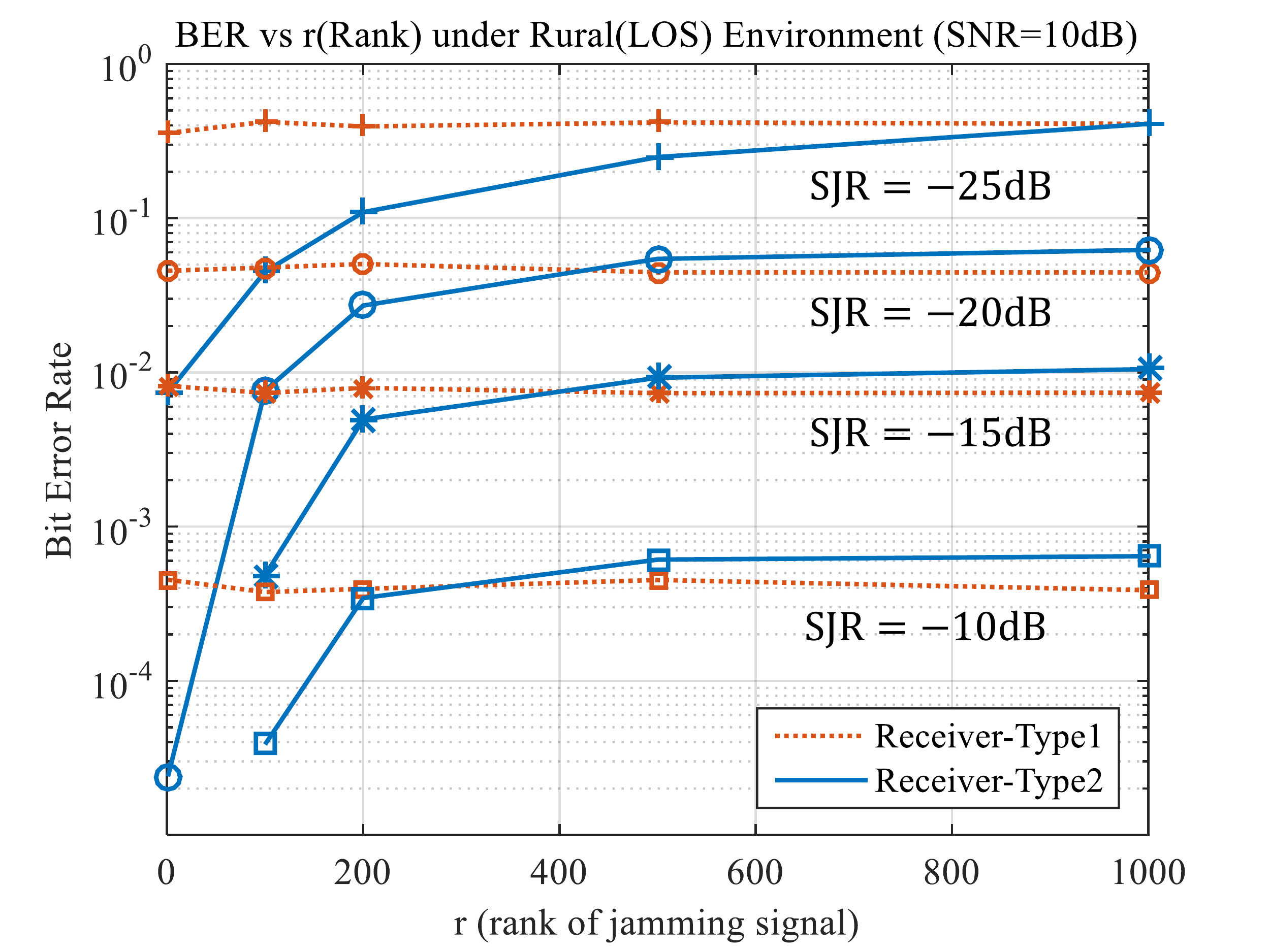}
	\caption{BER versus $r$ (rank of jamming signal) change with SNR fixed to 10dB under rural environments (LOS).}
\end{figure}

\begin{figure}[t]
	\centering
	\includegraphics[width = 8.6cm]{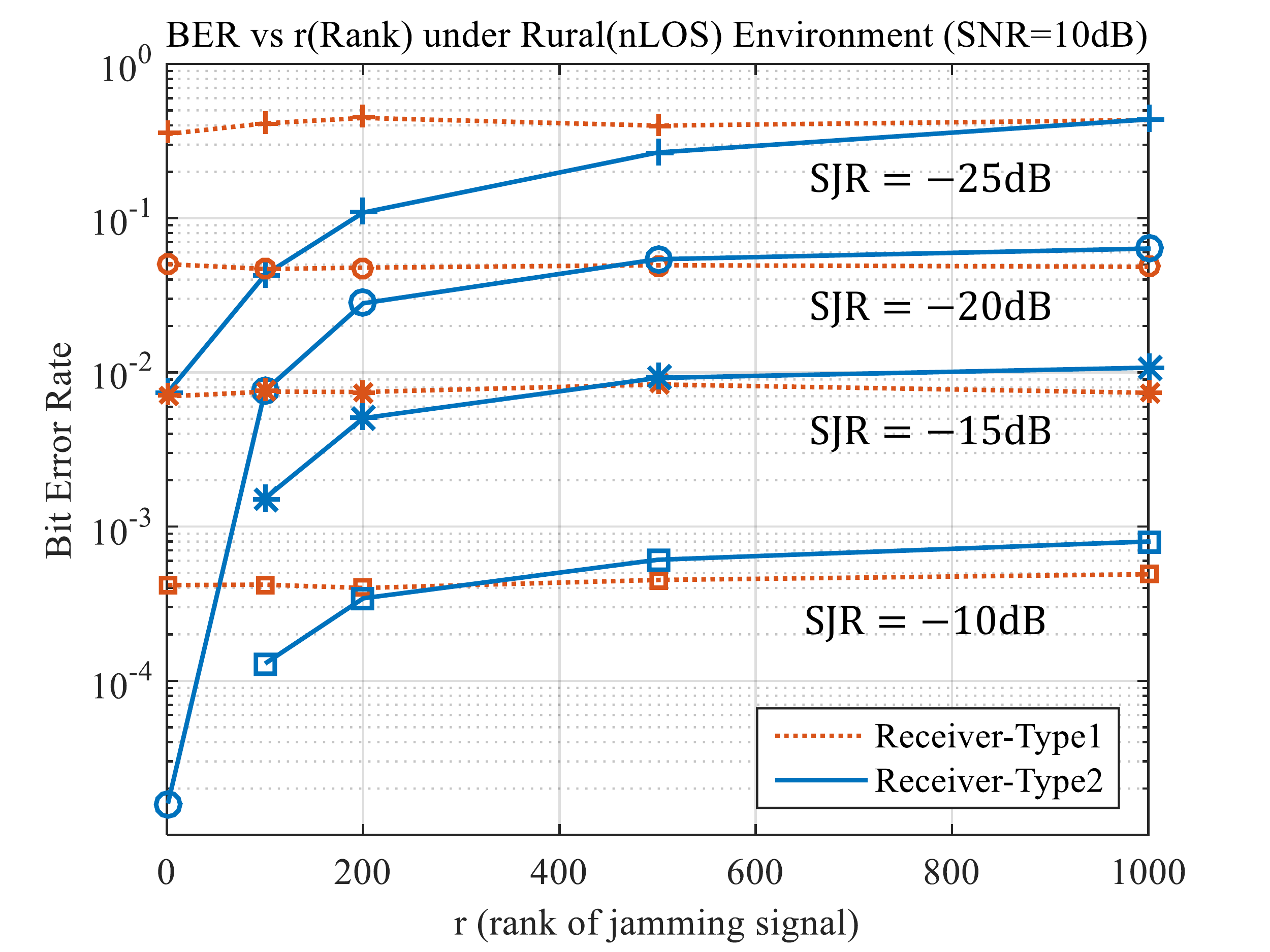}
	\caption{BER versus of $r$ (rank of jamming signal) change with SNR fixed to 10dB under rural environments (nLOS).}
\end{figure}

Fig. 3 presents the anti-jamming performance of the aforementioned two receivers versus SJR under the urban environments including a LOS path with $\textrm{SNR}=5 \textrm{ and } 10\, \textrm{dB}$ on the left and right figures, respectively. 
Each subfigure considers the MT-FH uplink jammer with different jamming ranks of 1, 100, and 1000.
The blue curves are for the Receiver-Type2, and the red curves are for the Receiver-Type1. 
The results show that the Receiver-Type2 outperforms the Receiver-Type1 in most cases and the BER performance of the Receiver-Type2 increases as rank decreases. 
Especially, it is noted that the Receiver-Type2 completely decomposes the transmitted DS-CDMA signal matrix and the MT-FH jamming signal matrix with the rank $r=1$, when the signal power is larger than an SJR level of -20dB with a fixed SNR level of 10dB.
This implies that the typical MT jamming without FH can be easily separated by the Robust PCA even with very high SJR value.
It is also noteworthy that typical uplink jammers in GPSs are commonly simple single tone pulse generators with a high power amplifier, which can be effectively mitigated by using the proposed receiver. 
In the case that MT-FH jamming signals hop every bit duration, whose the rank increases up to $r=1000$, the Receiver-Type2 for SNR=10dB still guarantees a comparable anti-jamming performance compared to its counterpart. 
In another case of SNR=5dB, although the Receiver-Type1 outperforms the Receiver-Type2 for $r=1000$, the Receiver-Type2 performs better for MT-FH jamming signals with low hopping rates.
Simulation results also remark that the BER results of the Receiver-Type1 are almost the same and independent with respect to the rank of the jamming signal for both SNR=5dB and 10dB. 
This implies that ICA does not utilize low-dimensionality to decompose the signals.

In Fig. 4, we simulate the BER of the two receivers in similar conditions of Fig.3 except that the urban environment with nLOS path is considered.
From the Fig. 4, we observe that the BER performance of the Receiver-Type2 increases when the jammer decreases its hopping rate (the rank $r$). 
However, it should be noticed that MT-FH jammers, which require a high-rank $r$, are not common due to their high complexity and hardware costs in practical satellite communication systems. 
The figure also shows that the BERs of the Receiver-Type1/2 are saturated to $1.5\cdot10^{-3}$ and $5\cdot10^{-6}$ as SJR increases when SNR=5dB and 10dB, respectively. 
This result is explained by the effects of the LMS channel under the urban environment with nLOS that implies a highly fading channel. 
Similar to Fig. 3, the worst BER performance of the Receiver-Type2 is observed when $r=1000$.

Fig. 5 and 6 plot the BERs of the Receiver-Type1/2 under the rural environment with LOS/nLOS.
The BERs of the Receiver-Type1 under the rural environments is almost equal to the BERs under the urban environments. One difference is that the BERs under the rural environment with nLOS are not saturated within the simulated SJR region.
For the case of $r=1$ (no hopping) of the Receiver-Type2, the BERs approach roughly $10^{-5}$ at SJR=-20dB and SNR=10dB while the BER of the Receiver-Type2 under the urban environment with LOS is $10^{-3}$.
One reason, why the BER of the Receiver-Type1/2 are not saturated and the Receiver-Type2 gives better AJ performance, is that the rural LMS channels are measured by fewer paths and long delay channel impulse responses compared to the urban environments.

\begin{figure*}[t!]
	\centering
	\subfloat[Runtimes and BER of the Receiver-Type1/2 versus the number of users $K$, with $M=128, N=100$]{
		\label{sfig:RunBERvsK}
		\centering
		\includegraphics[width = 5.6cm]{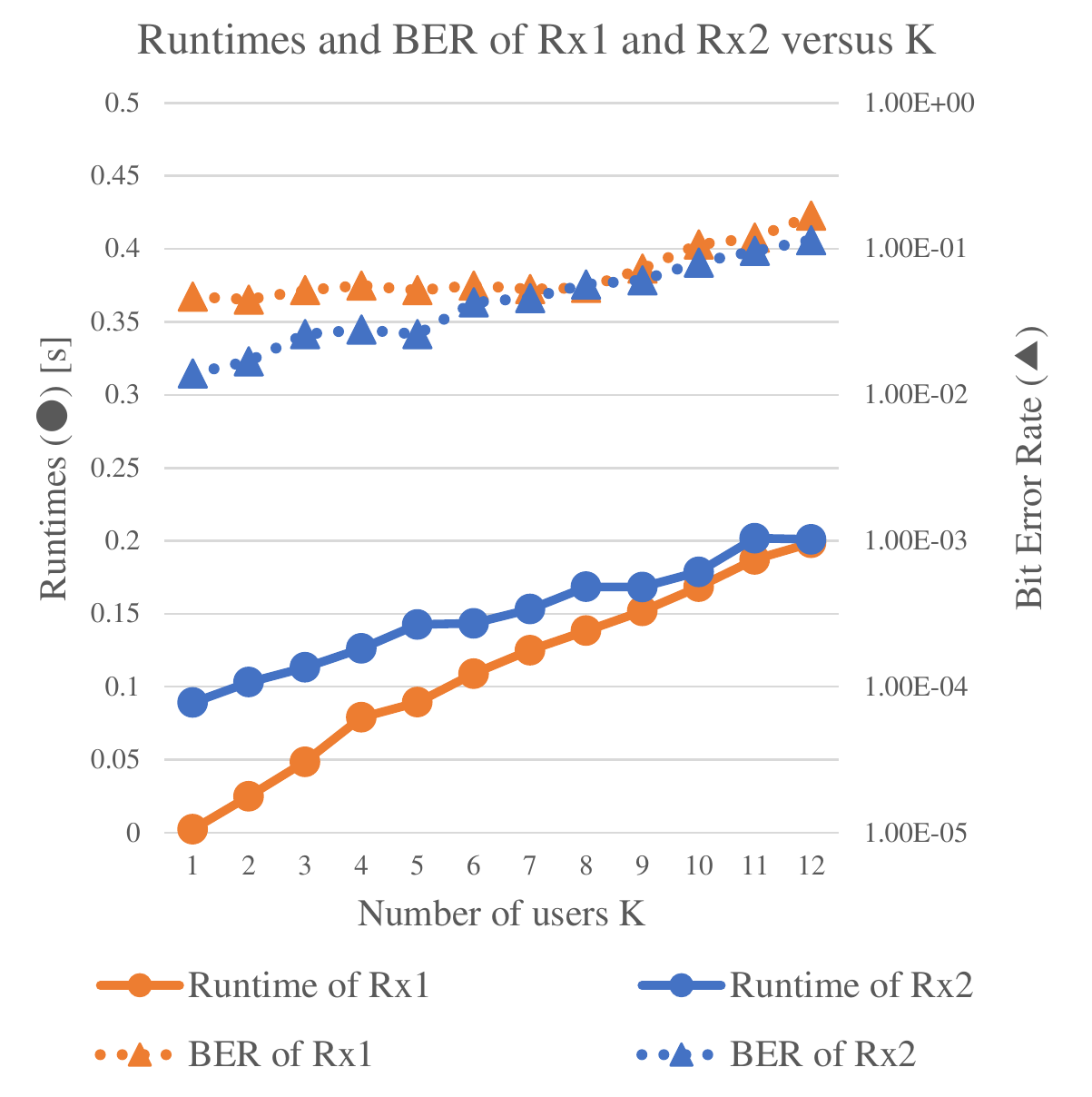}
	}
	\hfill
	\subfloat[Runtimes and BER of the Receiver-Type1/2 versus the spreading code length $M$ with $K=3, N=100$]{
		\label{sfig:RunBERvsM}
		\centering
		\includegraphics[width = 5.6cm]{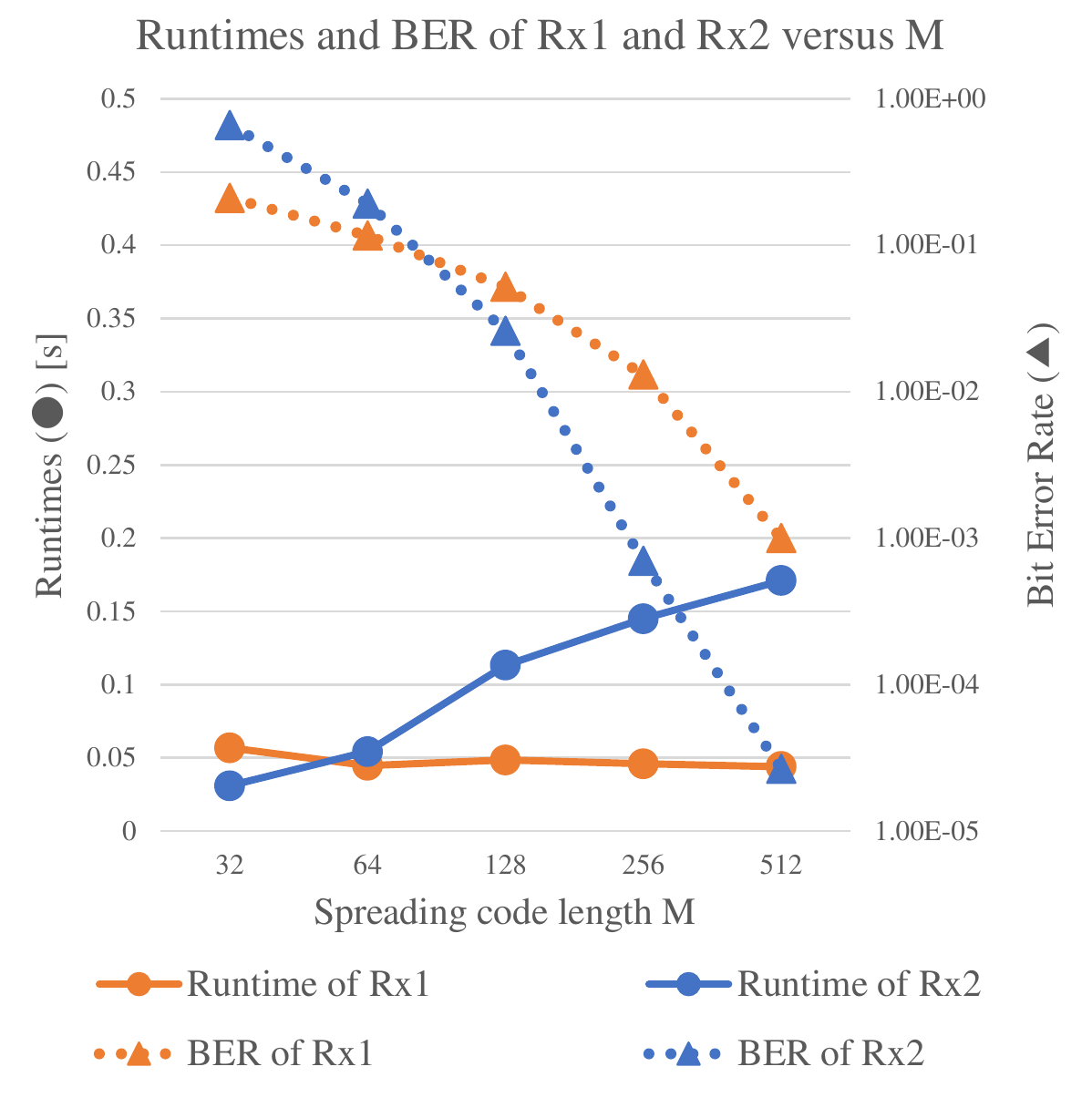}
	}
	\hfill
	\subfloat[Runtimes and BER of the Receiver-Type1/2 versus the number of bits $N$ with $K=6, M=128$]{
		\label{sfig:RunBERvsN}
		\centering
		\includegraphics[width = 5.6cm]{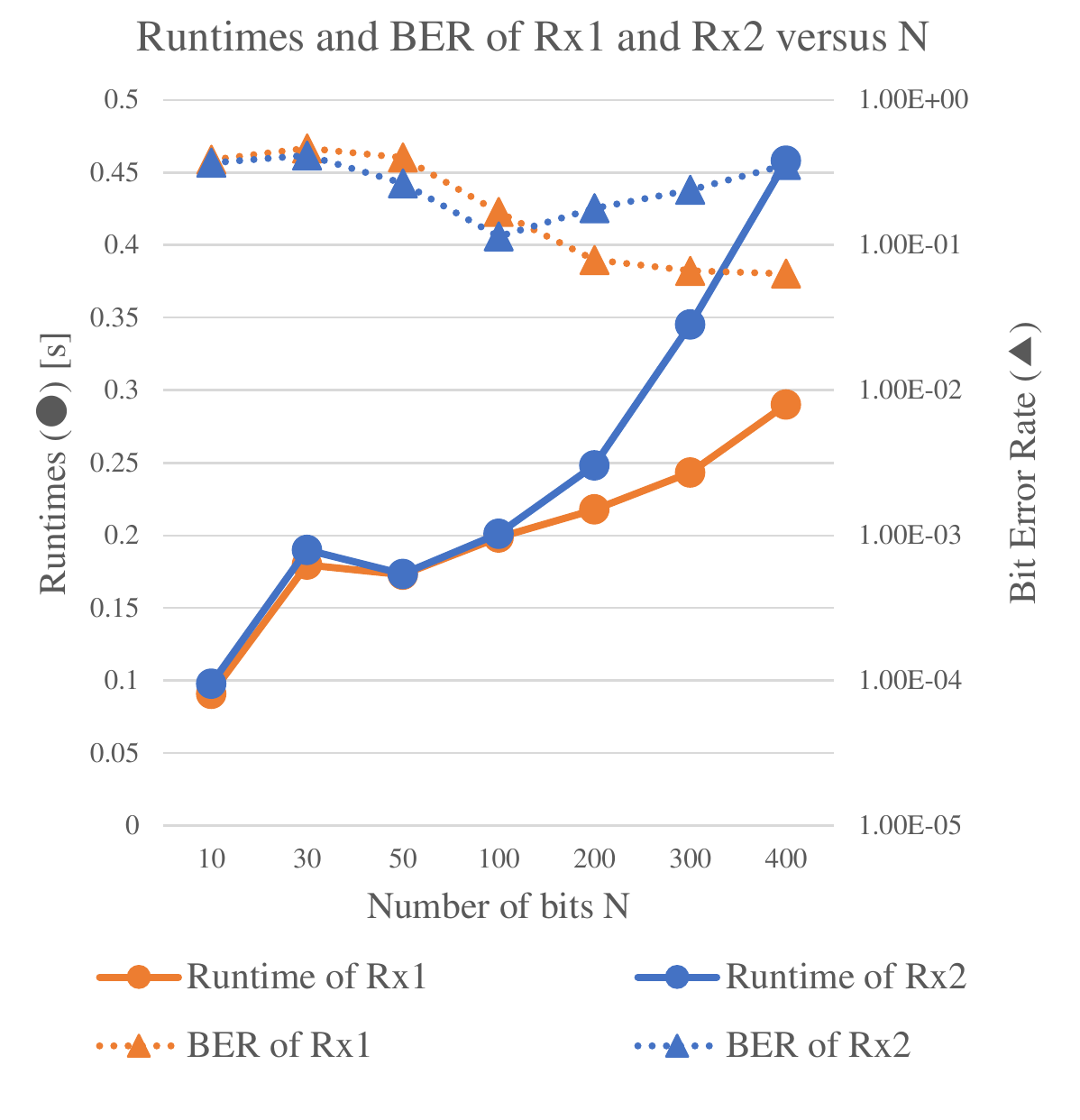}
	}
	\caption{Runtimes and BER of the Receiver-Type1/2 versus the number of users $K$, the spreading code length $M$, and the number of bits $N$. Simulation considers the rural environment with nLOS path, SNR=5dB, SJR=-10dB, and the jamming rank is $N/10$. }
	\label{fig:RunBERvsKMN}
\end{figure*}

In Fig. 7, 8, 9, and 10, we compare the anti-jamming performances of the Receiver-Type1/2 versus the rank of the jamming signal with $\textrm{SJR} = -25, -20, -15,\ \textrm{and} -10 \textrm{dB}$ under the MT-FH uplink jamming for four different environments.
Fig. 7 and 8 plot the BERs versus the rank-$r$ under urban environments with LOS/nLOS and Fig. 9 and 10 are under rural environments with LOS/nLOS.
Red dotted curves correspond to the BERs of the Receiver-Type1 for different SJRs, and blue curves are for the BERs of the Receiver-Type2. 
X-axis represents the rank of the jamming $r=(1, 100, 200, 500, 1000)$, which the minimum $r$ corresponds to no hopping and the maximum $r$ is for the case of hopping every bit duration.

Overall Fig. 7, 8, 9, and 10, as the rank of the jamming decreases, the anti-jamming capacity of the Receiver-Type2 increases. On the other hand, the Receiver-Type1 does not improve the BER performance although the rank of the jamming decreases. 
At a low-rank range of $r<200$, which represents less hopping MT-FH jammers, the Receiver-Type2 significantly outperforms the Receiver-Type1 for a wide range of SJRs. Moreover, at high-rank jammings, the Receiver-Type2 performs equally well or slightly worse than the Receiver-Type1 depending on SJR. 
The BER differences between the Receiver-Type1/2 for the high-rank jamming decrease as SJR decreases. In addition, ranges of rank values, where the Receiver-Type2 performs better than the Receiver-Type1, become wider as SJR decreases--in other words, the jamming power increases. 
It is observed that the range of rank values that Receiver-Type2 outperforms Receiver-Type1 is smaller when the LMS downlink channel becomes severe. 
Simulation results conclusively remark that the proposed Receiver-Type2 is more effective than the conventional Receiver-Type1 for low-rank ($r<200$), high power jammers, and less-severe multi-path environments. In addition, even for high-rank and more-severe multi-path channels, the Receiver-Type2 is competitive to the Receiver-Type1.

The CPU runtimes of the MATLAB implementations and BER performances of the Receiver-Type1 and the Receiver-Type2 with respect to various DS-CDMA system parameters are summarized in Fig. \ref{fig:RunBERvsKMN}. 
The subfigures for the number of users $K$, the spreading code length $M$, and the number of bits $N$ are presented in Fig. \ref{sfig:RunBERvsK}, Fig. \ref{sfig:RunBERvsM}, and Fig. \ref{sfig:RunBERvsN}, respectively. 
The rural environment with nLOS path is assumed, and SNR and the jamming rank are set to 5dB and $N/10$. 
In addition, BERs are measured at SJR of -10dB.
Generally, the results show that the computational time of the Receiver-Type2, combining Robust PCA and ICA, is comparable to that of the Receiver-Type1 using ICA only. 
The Fig. \ref{sfig:RunBERvsK} implies that the computational time of the Receiver-Type1 increases linearly as the number of users $K$ increases, while the gap between the runtime of the Receiver-Type2 and that of the Receiver-Type1 reduces. 
It is also seen increasing $K$ degrades the BER performances of both the Receiver-Type1 and the Receiver-Type2.
The Fig. \ref{sfig:RunBERvsM} shows that the spreading code length $M$ only linearly increases the CPU runtime of the Receiver-Type2, while the BER performance of the Receiver-Type2 is exponentially improving. 
Moreover, in the Fig. \ref{sfig:RunBERvsN}, we observe that the runtimes of both the Receiver-Type1 and the Receiver-Type2 increase as the number of bits $N$ increases. 
It is also noted that the additional time for combining Robust PCA algorithms on the Receiver-Type1, when the number of bits $N$ is less than 400, is less than the computational time of the Receiver-Type1 itself.

\section{Conclusion}

In this paper, we considered the anti-jamming problem of DS-CDMA receivers against the presence of uplink jammers under LMS communication systems.
We developed an anti-jamming DS-CDMA receiver that decomposes the received signal into the transmitted signal and the unintended uplink jamming signal by exploiting the fact that they are typically low-dimensionality.
Utilizing their low-dimensionality attributes, we suggested the integration of Robust PCA and ICA approaches, which are implemented by iALM and Fast ICA algorithms.

Anti-jamming performances of Receiver-Type1 (the conventional receiver using only ICA without Robust PCA) and Receiver-Type2 (the proposed receiver using both Robust PCA and ICA) were assessed in the scenarios that consider the MT-FH uplink jammer and practical downlink channels including urban and rural environments.
Simulation results show that Robust PCA in Receiver-Type2 achieves significant performance improvement as compared with the Receiver-Type1 for a wide range of the rank of the MT-FH jamming signal. This implies that Robust PCA separates various jamming signals more effectively than ICA only. The performance improvement increases as the rank decreases. 
For ranks lower than 200 that represent MT-FH jamming signals with less hopping, Receiver-Type2 outperforms Receiver-Type1. Even for large ranks that signify frequent hopping jamming, Receiver-Type2 shows a comparable performance to its counterpart.
In conclusion, our proposed receiver has potential applications in DS-CDMA based LMS systems under various uplink jammers.

\section*{Acknowledgment}
The authors gratefully acknowledge the support from Electronic Warfare Research Center (EWRC) at Gwangju Institute of Science and Technology (GIST), originally funded by Defense Acquisition Program Administration (DAPA) and Agency for Defense Development (ADD).



\begin{IEEEbiography}[{\includegraphics[width=1in,height=1.25in,clip,keepaspectratio]{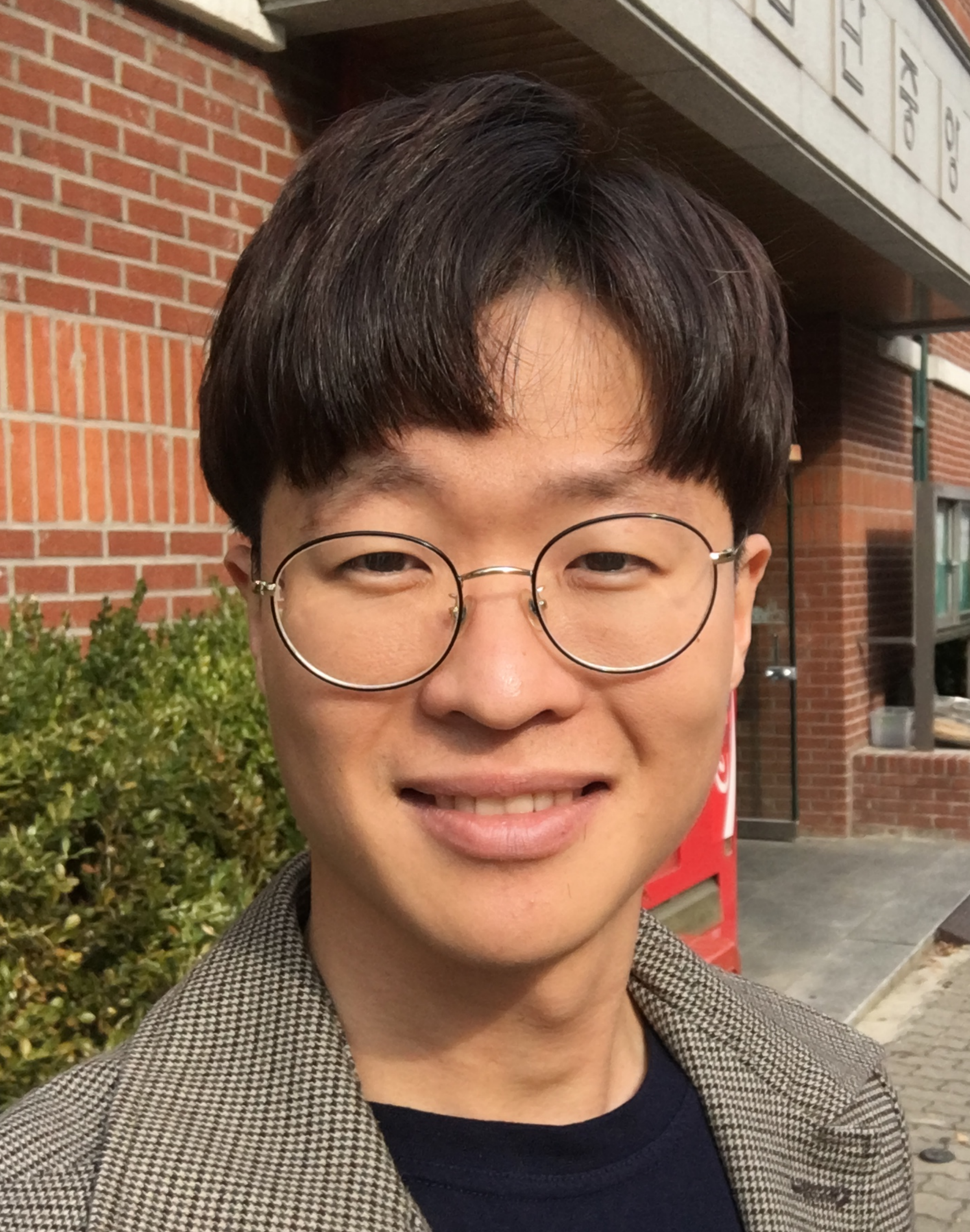}}]%
	{Hyoyoung Jung}
	 received the B.S. and M.S. degrees from Inha University and the Gwangju Institute of Science and Technology (GIST), South Korea, in 2011 and 2013, respectively. He is currently pursuing the Ph.D. degree
	with GIST. His research interests include statistical signal processing, machine learning, and anti-jamming communication systems.
\end{IEEEbiography}

\begin{IEEEbiography}[{\includegraphics[width=1in,height=1.25in,clip,keepaspectratio]{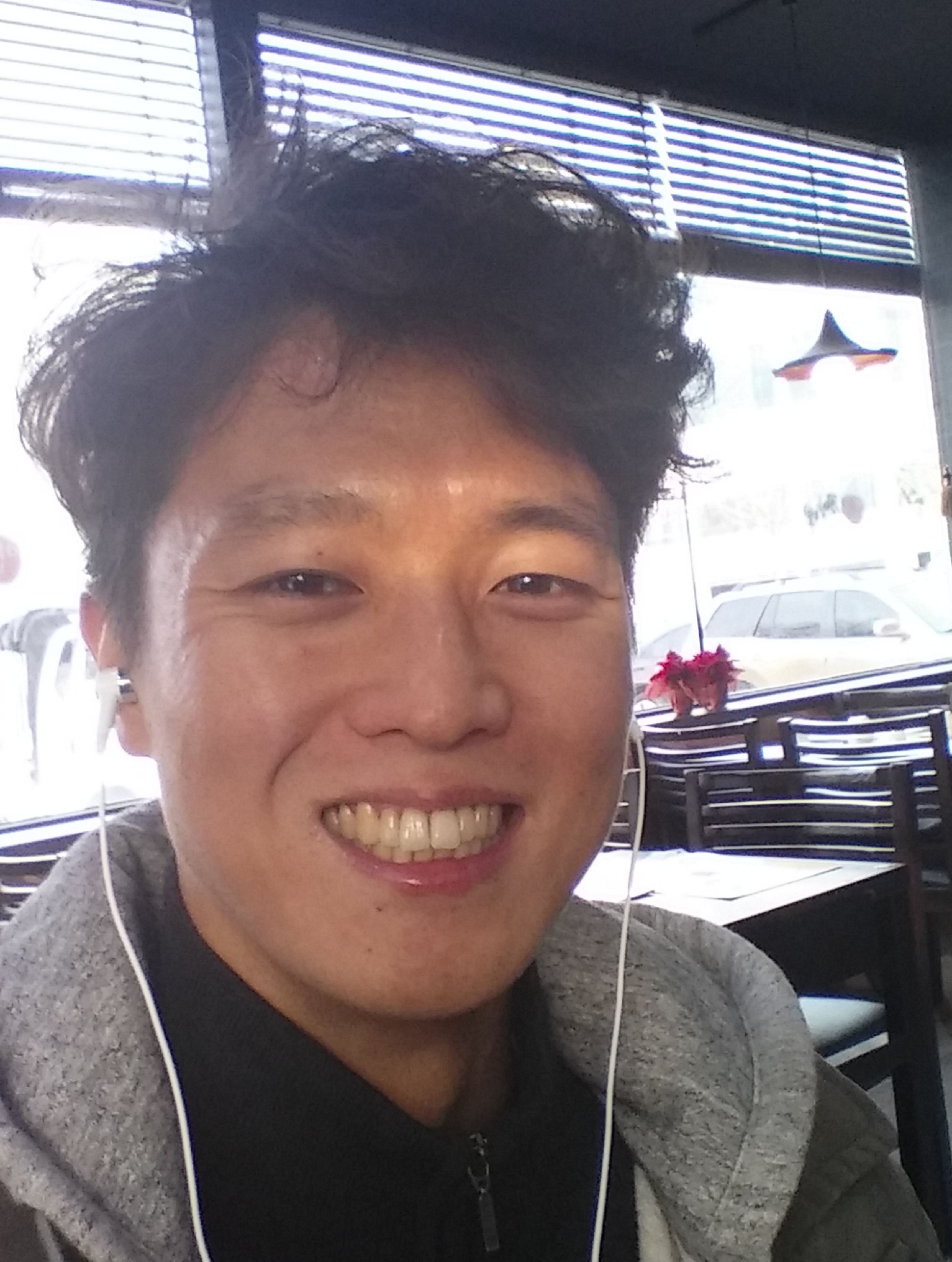}}]%
	{Jaewook Kang}
	received the B.S. degree in information and communications engineering (2009) from Konkuk University, South Korea, and the M.S. and Ph.D. degrees in information and communications engineering (2010,2015) from the Gwangju Institute of Science and Technology (GIST), South Korea. He is currently working in Soundlly Inc. as a signal and data processing engineer. His recent research interests include signal processing based on low-dimensionality, low-complex ultrasound receiver design, and machine learning techniques for IoT devices.
	
\end{IEEEbiography}
\begin{IEEEbiography}[{\includegraphics[width=1in,height=1.25in,clip,keepaspectratio]{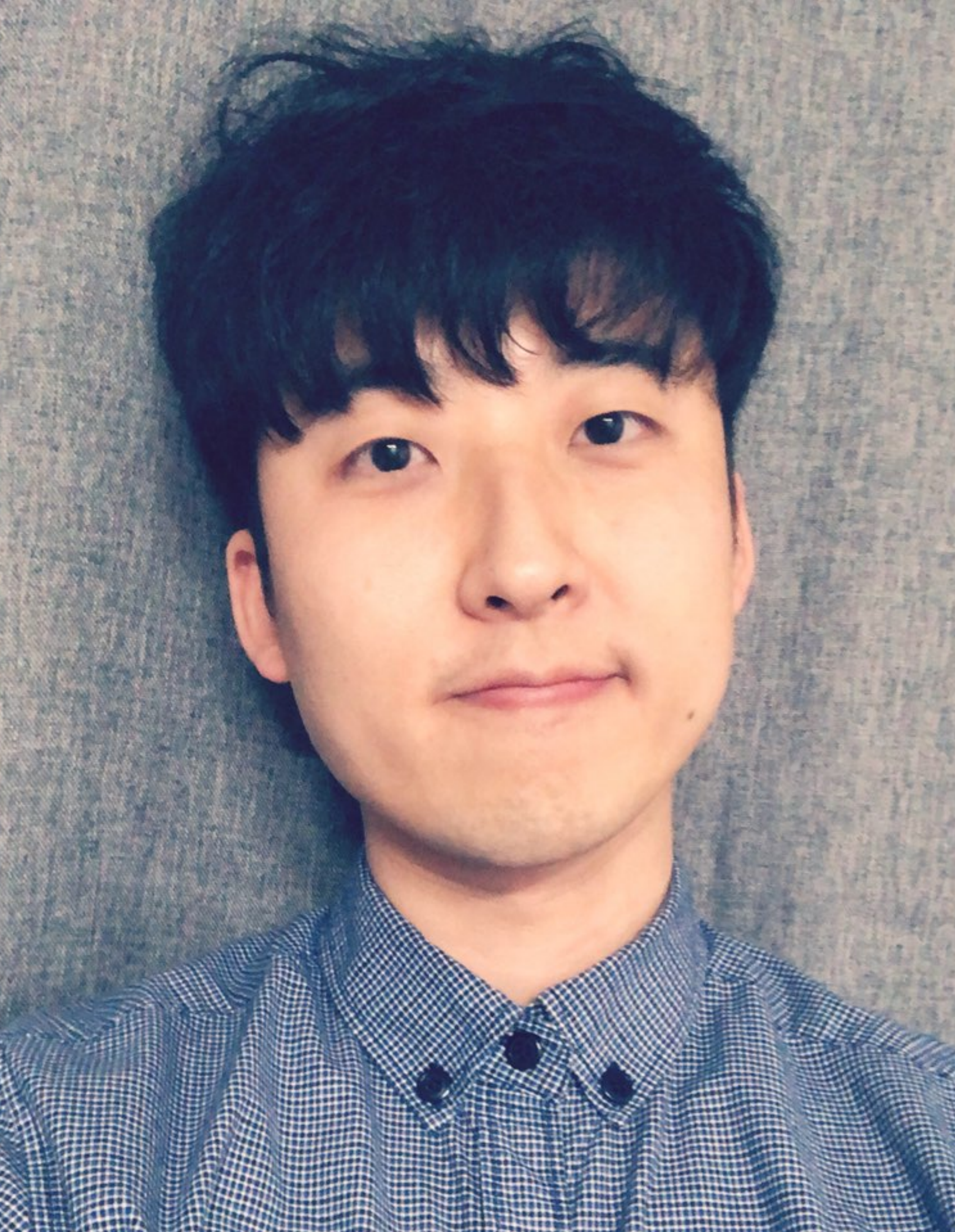}}]%
	{Tae Seok Lee}
	received the B.S. degree in control engineering from Kwangwoon University, South Korea, in 2015 and M.S. degree in electrical and computer engineering from GIST, South Korea, in 2017. He is currently working in Telecommunications Technology Association (TTA), as a senior research engineer. His research interests include the satellite communications and data processing under big data system.
\end{IEEEbiography}
\begin{IEEEbiography}[{\includegraphics[width=1in,height=1.25in,clip,keepaspectratio]{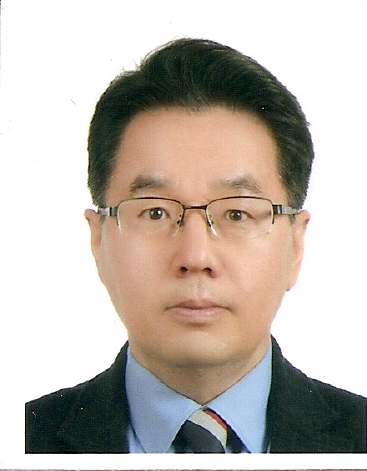}}]%
	{Suil Kim}
	received the B.S. and M.S. degrees in electrical engineering from Soongsil University, South Korea, in 1986 and 1988, respectively, and the Ph.D. degree in electrical engineering and computer science from the Korea Advanced Institute of Science and Technology, South Korea, in 2000. He is currently a principal researcher in the Agency for Defense Development (ADD) and a professor in the University of Science and Technology in Daejeon, South Korea. His research interests are on anti-jamming modems and interference cancellation for wireless communication systems.
\end{IEEEbiography}
\begin{IEEEbiography}[{\includegraphics[width=1in,height=1.25in,clip,keepaspectratio]{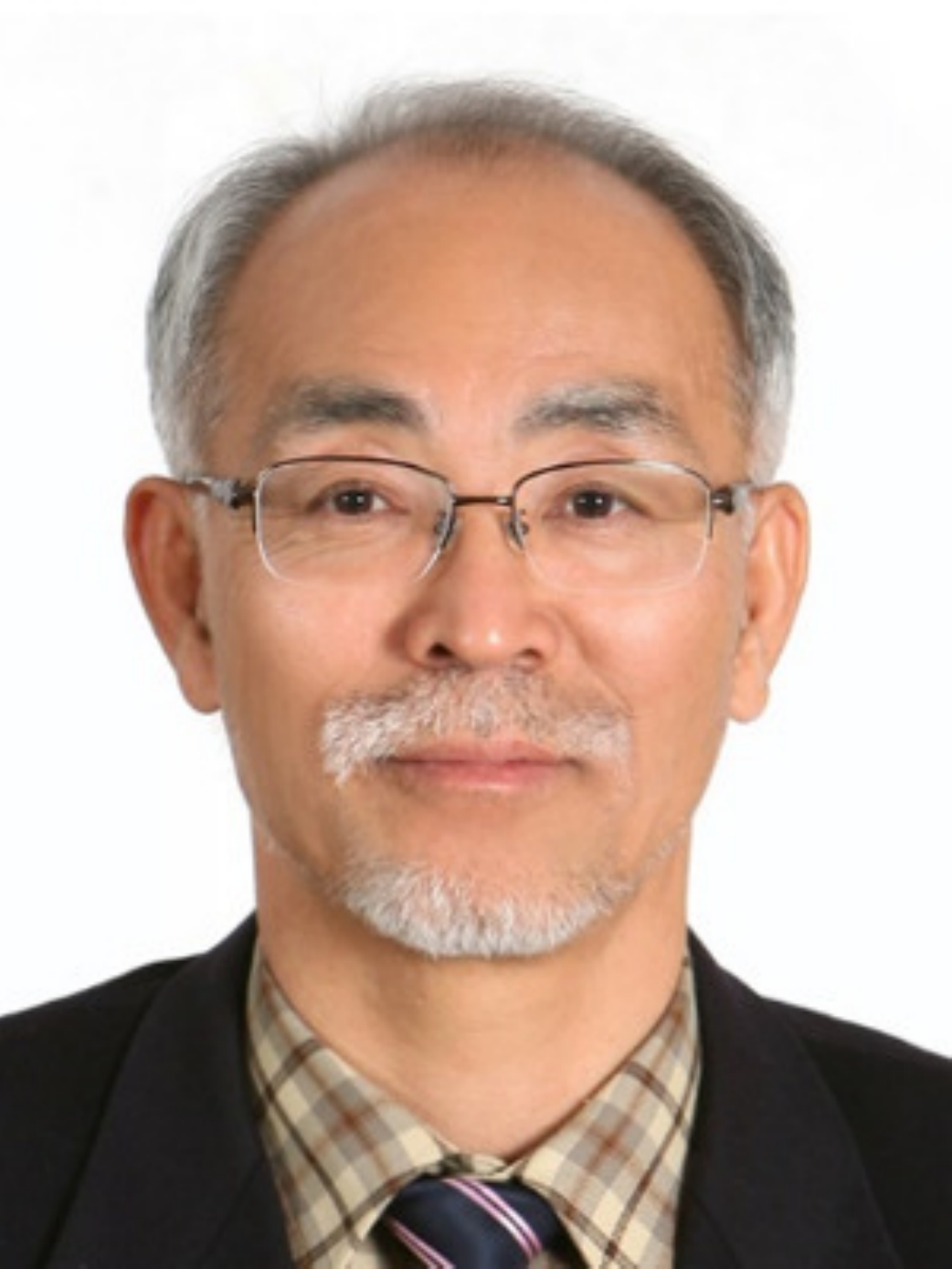}}]%
	{Kiseon Kim}
	received the B. Eng and M. Eng. Degrees in electronics engineering from Seoul National Unversity, Seoul, South Korea, in 1978 and 1980, respectively, and the Ph.D. degree in electrical engineering systems from the University of Southern California, Los Angeles, CA, USA, in 1987. From 1988 to 1991, he was with Schlumberger, Houston, TX, USA. From 1991 to 1994, he was with the Superconducting Super Collider Lab, TX, USA. In 1994, he joined Gwangju Institute of Science and Technology, Gwangju, South Korea, where he is currently a Professor. His current research interests include wideband digital communications system design, sensor network design, analysis and implementation both at the physical and at the resource management layer, and biomedical application design. Dr. Kim is the member of the National Academy of Engineering of Korea, the Fellow IET, and the Senior Editor of the IEEE Sensors Journal.
\end{IEEEbiography}

\end{document}